\documentclass[11pt,ams]{article}
\usepackage{graphicx}
\usepackage{amssymb}
\usepackage{amsmath}
\newcommand{\MSbar}{\overline{\mbox{MS}}}

\newcommand{\al}{\alpha}
\newcommand{\p}{\partial}
\newcommand{\oc}{\overline{c}}

\newcommand{\omu}{\overline{\mu}}
\newcommand{\lms}{\Lambda_{\overline{\mbox{\tiny{MS}}}}}
\newcommand{\Evac}{E_{\textrm{\tiny{vac}}}}

\newcommand{\wsigma}{\widetilde{\sigma}}

\newcommand{\og}{\overline{g}}
\newcommand{\om}{\overline{m}}

\newcommand{\sect}[1]{ \section{#1} \setcounter{equation}{0} }

\begin{document}
\date{}

\title{\textbf{A combined study of the gluon and ghost condensates }$\left\langle A_{\mu
}^{2}\right\rangle \;$\textbf{and $\left\langle
\varepsilon^{abc}\overline{c}^{b}c^{c}\right\rangle $ in Euclidean
$SU(2)$ Yang-Mills theory in the Landau gauge}}
\author{ \textbf{M.A.L. Capri$^a$\thanks{marcio@dft.if.uerj.br}} \ , \textbf{D. Dudal}$^{b}$\thanks{david.dudal@ugent.be}{\
}\footnote{Research Assistant of the Research Foundation - Flanders (FWO-Vlaanderen)} \ , \textbf{J.A. Gracey$^{c}$\thanks{jag@amtp.liv.ac.uk}} \ ,
\textbf{V.E.R. Lemes$^{a}$\thanks{vitor@dft.if.uerj.br}}  \ , \\\textbf{R.F. Sobreiro}$^{a}$\thanks{%
sobreiro@uerj.br}  \ , \textbf{S.P. Sorella}$^{a}$\thanks{%
sorella@uerj.br}{\ }{\ }\footnote{Work supported by FAPERJ, Funda{\c
c}{\~a}o de Amparo {\`a} Pesquisa do Estado do Rio de Janeiro, under
the program {\it Cientista do Nosso Estado}, E-26/151.947/2004.}  \
,  \textbf{H. Verschelde}$^{b}$\thanks{henri.verschelde@ugent.be} \\\\
\textit{$^{a}$\small{Departamento de F\'{\i }sica Te\'{o}rica}}\\
\textit{\small{Instituto de F\'{\i }sica, UERJ, Universidade do Estado do Rio de Janeiro}} \\
\textit{\small{Rua S{\~a}o Francisco Xavier 524, 20550-013 Maracan{\~a}}} \\
\textit{\small{Rio de Janeiro, Brasil}} \\[3mm]
\textit{$^{b}$\small{Ghent University}} \\
\textit{\small{Department of Mathematical Physics and Astronomy}} \\
\textit{\small{Krijgslaan 281-S9, B-9000 Gent, Belgium}}\\
[3mm] \textit{$^c$\small{Theoretical Physics Division} }\\
\textit{\small{Department of Mathematical Sciences}}\\
\textit{\small{University of Liverpool}}\\
\textit{\small{P.O. Box 147, Liverpool, L69 3BX, United Kingdom}} }
\maketitle

\begin{abstract}
\noindent The ghost condensate $\left\langle \varepsilon ^{abc}\overline{c}%
^{b}c^{c}\right\rangle $ is considered together with the gluon condensate $%
\left\langle A_{\mu }^{2}\right\rangle $ in $SU(2)$ Euclidean
Yang-Mills theories quantized in the Landau gauge. The vacuum
polarization ceases to be transverse due to the nonvanishing
condensate
$\left\langle \varepsilon ^{abc}\overline{c}%
^{b}c^{c}\right\rangle $. The gluon propagator itself remains
transverse. By polarization effects, this ghost condensate induces
then a splitting in the gluon mass parameter, which is dynamically
generated through $\left\langle A_{\mu }^{2}\right\rangle$. The
obtained effective masses are real when $\left\langle
A_\mu^2\right\rangle$ is included in the analysis. In the absence of
$\left\langle A_\mu^2\right\rangle$, the already known result that
the ghost condensate induces effective tachyonic masses is
recovered. At the one-loop level, we find that the effective
diagonal mass becomes smaller than the off-diagonal one. This might
serve as an indication for some kind of Abelian dominance in the
Landau gauge, similar to what happens in the maximal Abelian gauge.
\end{abstract}
\vspace{-20cm} \hfill LTH--660 \vspace{18cm}
\newpage

\sect{Introduction.} Vacuum expectation values of composite
operators, commonly known as vacuum condensates, play an important
role in quantum field theory. One can employ them to parametrize
certain nonperturbative effects. In the context of gauge theories,
the gluon condensate $\left\langle F_{\mu\nu}^2\right\rangle$ and
quark condensate $\left\langle \overline{q}q\right\rangle$ are
renowned examples \cite{Shifman:1978bx}.

\noindent In the last few years, there has been a growing interest
in condensates of dimension two. Most attention was paid to the
gluon condensate $\left\langle A_\mu^2\right\rangle$ in case of the
Landau gauge. We do not intend to give a complete overview of the
existing research, we refer to the papers
\cite{Gubarev:2000eu,Gubarev:2000nz,Boucaud:2001st,Boucaud:2005rm,Kondo:2001nq,Verschelde:2001ia,Dudal:2002pq,
Dudal:2003vv,Dudal:2003by,Dudal:2004rx,Browne:2004mk,Gracey:2004bk,Dudal:2005na,Furui:2005bu,Gubarev:2005it,Slavnov:2005av,Suzuki:2004dw,Suzuki:2004uz}
and references therein, covering theoretical, phenomenological,
lattice and computational topics concerning the mass dimension two
gluon condensate. We have studied this condensate and its
generalizations to other gauges, such as the linear covariant, the
Curci-Ferrari, and the maximal Abelian gauges. In particular, we
developed the so-called LCO formalism, allowing us to construct a
renormalizable effective potential obeying a homogenous
renormalization group equation for a Local Composite Operator like
$A_\mu^2$, see e.g.
\cite{Verschelde:2001ia,Dudal:2002pq,Dudal:2003by,Dudal:2004rx}. The
renormalizability properties and relations between various
renormalization group functions can be proven to all orders of
perturbation theory by making use of the algebraic renormalization
technique \cite{Piguet:1995er}. According to the LCO construction,
an effective tree level gluon mass is dynamically generated due to
$\left\langle A_\mu^2\right\rangle\neq0$.

\noindent Perhaps less known is the concept of the ghost condensates
like $\left\langle f^{abc}\oc^b c^c\right\rangle$, $\left\langle
f^{abc}c^b c^c\right\rangle$ and $\left\langle f^{abc}\oc^b
\oc^c\right\rangle$. For the benefit of the reader, let us provide
here a short overview.

\noindent The ghost condensate $\left\langle \varepsilon^{3bc}\oc^b
c^c\right\rangle$ was first introduced in the maximal Abelian gauge
(MAG) in
\cite{Schaden:1999ew,Kondo:2000ey,Schaden:2000fv,Schaden:2001xu} in
case of the gauge group $SU(2)$. The MAG is a partial nonlinear
gauge fixing which is useful for the dual superconductivity picture
of low energy QCD. Due to the nonlinearity of the MAG, a quartic
ghost interaction needs to be introduced in the action for
renormalizability purposes \cite{Min:1985bx,Fazio:2001rm}. This
four-ghost interaction was decomposed by means of an auxiliary field
$\sigma$, and a one-loop effective potential for the ghost
condensate $\left\langle \varepsilon^{3bc}\oc^b
c^c\right\rangle\sim\left\langle\sigma\right\rangle$ was calculated.
A nonzero vacuum expectation value $\left\langle\sigma
\right\rangle$ is favoured as it lowers the vacuum energy. It was
consequently used to construct an effective mass for the
off-diagonal gluons, at one-loop order. The diagonal gluons remained
massless. This result was interpreted as analytical evidence for the
Abelian dominance hypothesis \cite{Kondo:2000ey}, according to which
the low energy regime of QCD should be expressed solely in terms of
Abelian degrees of freedom \cite{Ezawa:1982bf}. Lattice evidence of
this Abelian dominance in case of the MAG was presented in
\cite{Suzuki:1989gp,Suzuki:1992gz,Hioki:1991ai}. To our knowledge,
there is no analytical proof of the Abelian dominance. An argument
that can be interpreted in favour of it, is the fact that the
off-diagonal gluons would acquire a mass through a dynamical
mechanism. At energies below the scale set by this mass, the
off-diagonal gluons should decouple, and in this way one should end
up with an Abelian theory at low energies. It is worth noticing that
lattice simulations of the $SU(2)$ MAG revealed an off-diagonal mass
of approximately $1.2$ GeV \cite{Amemiya:1998jz,Bornyakov:2003ee},
while the diagonal gluons behaved masslessly \cite{Amemiya:1998jz}
or at least almost masslessly \cite{Bornyakov:2003ee}.

\noindent Returning to the ghost condensation in the MAG, it was
shown in \cite{Dudal:2002xe,Sawayanagi:2003dc} that, contrary to the
claim in
\cite{Schaden:1999ew,Kondo:2000ey,Schaden:2000fv,Schaden:2001xu},
the induced off-diagonal mass induced was tachyonic, at least at the
considered one-loop order. As such, it could not be taken as
analytical evidence for the Abelian dominance. Another condensate,
namely\footnote{$\al$ is the MAG gauge parameter, while the color
index $\beta$ runs only over the $N(N-1)$ off-diagonal generators of
$SU(N)$.} $\left\langle \frac{1}{2}A_\mu^\beta
A_\mu^\beta+\al\oc^\beta c^\beta\right\rangle$ that could be
responsible for a real-valued off-diagonal gluon mass was proposed
in \cite{Kondo:2001nq} and investigated thoroughly in
\cite{Dudal:2004rx} using the LCO formalism.

\noindent In \cite{Dudal:2002xe}, it was signalled that the
effective potential, obtained using the decomposition of the
four-point ghost interaction can cause renormalization problems
beyond one-loop order and that the LCO formalism would be more
suitable to discuss the ghost condensation. A further aspect of the
ghost condensation was pointed out in \cite{Lemes:2002ey}, where it
was shown that an alternative decomposition of the quartic ghost
interaction led to the two Faddeev-Popov charged ghost condensates
$\left\langle \varepsilon^{3bc}c^b c^c\right\rangle$ and
$\left\langle \varepsilon^{3bc}\oc^b \oc^c\right\rangle$, instead of
$\left\langle \varepsilon^{3bc}\oc^b c^c\right\rangle$. This should
be not too big a surprise, as the ghost condensation is an order
parameter for a continuous $SL(2,\mathbb{R})$ symmetry present in
the MAG \cite{Schaden:1999ew,Kondo:2000ey,Dudal:2002ye}. Said
otherwise, a nonvanishing ghost condensate like $\left\langle
\varepsilon^{3bc}\oc^b c^c\right\rangle$ induces a breakdown of the
$SL(2,\mathbb{R})$ symmetry. It turns out in fact that the
$SL(2,\mathbb{R})$ rotations interchange the different channels,
i.e. $\left\langle \varepsilon^{3bc}\oc^b c^c\right\rangle$,
$\left\langle \varepsilon^{3bc}c^b c^c\right\rangle$, $\left\langle
\varepsilon^{3bc}\oc^b \oc^c\right\rangle$,  in which the ghost
condensation might occur.

\noindent Later on, the ghost condensation was discussed in the case
of the Curci-Ferrari gauge \cite{Lemes:2002jv}, which also possesses
the $SL(2,\mathbb{R})$ invariance \cite{Dudal:2002ye}. This was
achieved by decomposing the four-ghost interaction which is present
for nonvanishing gauge parameter. This brings one to the Landau
gauge, which corresponds to the Curci-Ferrari gauge with vanishing
gauge parameter. As there is no longer an interaction to be
decomposed, it is less clear how to construct an effective potential
for the ghost condensates in this case. However, in
\cite{Lemes:2002rc}, it was shown that the LCO formalism does allow
to construct an effective potential for the ghost condensates
$\left\langle f^{abc}c^b c^c\right\rangle$ and $\left\langle
f^{abc}\oc^b \oc^c\right\rangle$ in the Landau gauge. This study was
pursued in \cite{Dudal:2003dp}, where it has been proven that the
operators $f^{abc}\oc^b c^c$, $f^{abc}c^b c^c$ and $f^{abc}\oc^b
\oc^c$ can be simultaneously coupled to the Yang-Mills action, while
preserving the $SL(2,\mathbb{R})$ symmetry, using the LCO setup. It
was consequently shown that the condensation\footnote{$\left\langle
f^{abc}\oc^b c^c\right\rangle$ was called the Overhauser condensate,
while $\left\langle f^{abc}c^b c^c\right\rangle$ and $\left\langle
f^{abc}\oc^b \oc^c\right\rangle$ the BCS condensates. This
nomenclature was based upon a similar kind of phenomenon happening
in the theory of superconductivity.} can occur in different
channels, i.e. $\left\langle f^{abc}\oc^b c^c\right\rangle$ and
$\left\langle f^{abc}c^b c^c\right\rangle$, $\left\langle
f^{abc}\oc^b \oc^c\right\rangle$. However, the corresponding vacua
are equivalent, being connected trough rotations of the broken
symmetry. Details on the symmetry breaking and the construction of
the potential can be found in \cite{Dudal:2003dp}. As the ghost
condensates carry a color index, let us specifically mention that we
have given an argument that the apparent breaking of the global
color symmetry should not be observed in the physical sector of the
theory \cite{Dudal:2003dp}. One can also argue that the Goldstone
particles associated to the broken continuous $SL(2,\mathbb{R})$
symmetry are unphysical \cite{Schaden:1999ew,Dudal:2003dp}.

\noindent Although the concept of a ghost condensate like
$\left\langle \varepsilon^{abc}\oc^b c^c\right\rangle$ might seem
unusual, it has many features in common with the fermion
condensation occurring in models with a four-fermion interaction as,
for example, the Gross-Neveu model. Considering the MAG, the ghost
condensation and the induced symmetry breaking can be directly
compared to what happens in the Gross-Neveu model
\cite{Gross:1974jv} or in other models with a quartic interaction.
Decomposing in fact the quartic interaction via an auxiliary field
allows us to construct an effective potential, and to analyze the
existence of a possible condensation and the related symmetry
breaking. The original setup of the ghost condensation, as it was
discussed in
\cite{Schaden:1999ew,Kondo:2000ey,Dudal:2002xe,Lemes:2002ey}, is
essentially not much different from the analysis of the Gross-Neveu
model. Let us also mention that the LCO formalism was first
developed to construct a meaningful effective potential for the
Gross-Neveu model for any number of fermions at any order of
perturbation theory \cite{Verschelde:1995jj}.

\noindent In this work, we shall continue our study on the gluon and
ghost condensates for the gauge group $SU(2)$ in case of the Landau
gauge. For the first time, we present a combined study of both
condensates, namely $\left\langle A_\mu^2\right\rangle$ and
$\left\langle \varepsilon^{abc}\oc^b c^c\right\rangle$. In Section
2, we shall discuss the renormalizability issues using the algebraic
renormalization. Section 3 contains a summary of the LCO formalism,
the calculation of the one-loop effective potential and the
determination of the vacuum configuration. Hereafter, we discuss the
consequences of a nonvanishing gluon and ghost condensate. We shall
show that the vacuum polarization is no longer transverse, the
breaking being directly proportional to the ghost condensate.
Moreover, we shall prove that the gluon propagator itself remains
transverse. These results will be first discussed by deriving the
Slavnov-Taylor identities in the condensed vacuum. Then, we shall
illustrate them with explicit one-loop calculations. These issues
will be handled in Sections 4, 5 and 6. In Section 7, we find
another interesting consequence of the ghost condensate. Due to
polarization effects on the gluon propagator, the effective
dynamical gluon mass generated through $\left\langle
A_\mu^2\right\rangle$ undergoes a shift which differs for the
diagonal and off-diagonal gluons. More precisely, it is found that
the effective diagonal gluon mass is smaller than the off-diagonal
one. However, unlike the results obtained in the absence of the
gluon condensate $\left\langle A_\mu^2\right\rangle$
\cite{Dudal:2002xe,Sawayanagi:2003dc}, both masses remain now real.

\noindent Similarly to what happens in the MAG
\cite{Amemiya:1998jz,Kondo:2000ey,Bornyakov:2003ee,Dudal:2004rx},
the fact that the off-diagonal mass is larger than the diagonal one,
could be interpreted as evidence that a kind of Abelian dominance
might also take place in the Landau gauge.

\sect{Renormalizability of the LCO formalism incorporating both
gluon and ghost condensates.} In this section, we shall prove that
the simultaneous introduction of the composite operators
$A^{2}_{\mu}$ and $gf^{abc}\bar{c}^{b}c^{c}$ allows for a
multiplicatively renormalizable action, in the Landau gauge.

\noindent We shall work in Euclidean space time. The Yang-Mills
action in the Landau gauge, $\partial_{\mu}A^{a}_{\mu}=0$, reads
\begin{eqnarray}
S&=& S_{\mathrm{YM}}+S_{\mathrm
{GF}}\;,\nonumber\\
S_{\mathrm{YM}} &=& \int d^4\!x\,\biggl(%
\frac{1}{4}F^{a}_{\mu\nu}F^{a}_{\mu\nu}\biggr)\;,\nonumber \\
S_{\mathrm{GF}} &=&s \int d^4\!x\,\Bigl(%
\bar{c}^a\partial_{\mu}A^{a}_{\mu}\Bigr)=\int
d^4\!x\,\Bigl(b^{a}\partial_{\mu}A^{a}_{\mu}+\bar{c}^{a}\partial_{\mu}D_{%
\mu}^{ab}c^{b}\Bigr)\;,
\label{YM&GF}
\end{eqnarray}
\noindent where $D_{\mu}^{ab}$ is the adjoint covariant derivative
which is defined by,
\begin{equation}
D_{\mu}^{ab}=\delta^{ab}\partial_{\mu}-gf^{abc}A^{c}_{\mu}\;.
\label{cov.derivative}
\end{equation}
\noindent The action (\ref{YM&GF}) enjoys the nilpotent BRST
symmetry
\begin{eqnarray}
sA^{a}_{\mu}&=&-D_{\mu}^{ab} c^{b}\;,\nonumber\\
sc^{a}&=&\frac{g}{2}f^{abc}c^{b}c^{c}\;,\nonumber\\
s\bar{c}^{a}&=&b^{a}\;,\nonumber\\
sb^a &=&0\;,
\end{eqnarray}
\noindent with
\begin{eqnarray}
sS&=&0\nonumber\;,\\ s^2&=&0\;.
\end{eqnarray}
\noindent We introduce two BRST doublets of sources
\begin{eqnarray}
s\tau&=&J \;,\;\;\;\;\;sJ=0\;, \\
s\lambda^{a}&=&\omega^{a}\;,\;\;\; s\omega^{a}=0\;,
\label{doublets}
\end{eqnarray}
allowing us to couple the composite operators $%
A^{2}_{\mu}$ and $gf^{abc}\bar{c}^{b}c^{c}$ to the action
(\ref{YM&GF}) in a BRST invariant fashion
\begin{eqnarray}
S^{\prime}&=&S_{\mathrm{ YM}}+S_{\mathrm{GF}}+s\int d^4\!x\,\biggl(\frac{1}{2}\tau A^{a}_{\mu}A^{a}_{\mu}+%
\frac{\zeta}{2}\tau J+gf^{abc}\lambda^a\bar{c}^{b}c^{c}-\frac{\rho}{2}%
\omega^{a}\lambda^{a}\biggr)\nonumber\\
&=&\int d^4\!x\,\biggl(\frac{1}{4}%
F^{a}_{\mu\nu}F^{a}_{\mu\nu}+b^{a}\partial_{\mu}A^{a}_{\mu}+\bar{c}%
^{a}\partial_{\mu}D_{\mu}^{ab}c^{b}+gf^{abc}\omega^{a}\bar{c}%
^{b}c^{c}-gf^{abc}\lambda^{a}b^{c}c^{c}\nonumber\\&+&\frac{g^2}{2}f^{abc}f^{cde}\lambda^{a}\bar{c}^{b}c^{d}c^{e}-
\frac{\rho}{2}\omega^{a}\omega^{a}+\frac{1}{2}JA^{a}_{\mu}A^{a}_{\mu}+\tau
A^{a}_{\mu}\partial_{\mu}c^{a}-\frac{\zeta}{2}J^{2}\biggr)\;.
\label{YM+GF+LCO}
\end{eqnarray}
The terms quadratic in the sources $\omega^a$ and $J$ are allowed by
power counting and are needed to remove the novel divergences
appearing in the vacuum correlators $\left\langle \left(
f^{abc}\oc^a c^b\right)(x)\left( f^{klm}\oc^l c^m\right)(y)
\right\rangle$ and $\left\langle A_\mu^2(x)A_\mu^2(y)\right\rangle$
for $x\rightarrow y$. $\rho$ and $\zeta$ are called the LCO
parameters.

\noindent For completeness, we also have to introduce external
sources for the BRST variations of the elementary fields
$A^{a}_{\mu}$ and $c^{a}$:
\begin{equation}
S_{\mathrm{ext}}=\int d^4\!x\,\biggl(-\Omega^{a}_{\mu}D_{\mu}^{ab}c^{b}+%
\frac{g}{2}f^{abc}L^{a}c^{b}c^{c}\biggr) \;, \label{ext}
\end{equation}
\noindent with
\begin{equation}
s\Omega^{a}_{\mu}=0,\qquad sL^{a}=0.
\end{equation}
\noindent The mass dimension and the ghost number of the fields and
sources are listed in Table \ref{table1}.

\begin{table}[t]
\centering
\begin{tabular}{|c|c|c|c|c|c|c|c|c|c|c|}
\hline
& $A^{a}_{\mu}$ & $c^{a}$ & $\bar{c}^{a}$ & $b^{a}$ & $\tau$ & $J$ & $%
\Omega^{a}_{\mu}$ & $L^{a}$ & $\lambda^{a}$ & $\omega^{a}$ \\ \hline
dimension & 1 & 0 & 2 & 2 & 2 & 2 & 3 & 4 & 2 & 2 \\
ghost number & 0 & 1 & $-1$ & 0 & $-1$ & 0 & $-1$ & $-2$ & $-1$ & 0
\\ \hline
\end{tabular}
\caption{Quantum numbers of the field and sources} \label{table1}
\end{table}
\noindent In principle, the action (\ref{YM+GF+LCO}) can be
supplemented with extra terms in the sources which are allowed by
power counting,
\begin{eqnarray}
S^{\prime}_{\mathrm{ext}} &=& s\int d^4\!x\,%
\biggl(\beta\frac{g}{2}f^{abc}\lambda^{a}\lambda^{b}c^s+\gamma\lambda^{a}%
\partial_{\mu}A^{a}_{\mu}\biggr)\nonumber\\
&=& \int d^4\!x\,\biggl(\beta
gf^{abc}\omega^{a}\lambda^{b}c^{c}+\beta\frac{g^{2}}{4}f^{abc}f^{cde}%
\lambda^{a}\lambda^{b}c^{d}c^{e}+\gamma\omega^{a}\partial_{\mu}A^{a}_{\mu}+%
\gamma\lambda^{a}\partial_{\mu}D_{\mu}^{ab}c^{b}\biggr)\;,
\label{supplement}
\end{eqnarray}
\noindent where $\beta$ and $\gamma$ are new, independent,
dimensionless couplings. The $S^{\prime}_{\mbox{\tiny ext}}$-part of
the action has to be introduced to actually prove the
renormalizability. This is our next task.

\noindent The complete action,
\begin{equation}
\Sigma=S^{\prime}+S_{\mathrm{ext}}+S^{\prime}_{\mathrm{ ext}}\;,
\label{sigma}
\end{equation}
\noindent obeys a few Ward identities.

\begin{itemize}
\item  {The Slavnov-Taylor identity
\begin{equation}
\mathcal{S}(\Sigma )=\int d^{4}\!x\,\biggl(\frac{\delta \Sigma
}{\delta \Omega _{\mu }^{a}}\frac{\delta \Sigma }{\delta A_{\mu
}^{a}}+\frac{\delta \Sigma }{\delta L^{a}}\frac{\delta \Sigma
}{\delta c^{a}}+b^{a}\frac{\delta \Sigma }{\delta
\bar{c}^{a}}+\omega ^{a}\frac{\delta \Sigma }{\delta \lambda
^{a}}+J\frac{\delta \Sigma }{\delta \tau }\biggr)=0\;.  \label{ST}
\end{equation}
}

\item  {The modified Landau gauge condition
\begin{equation}
\frac{\delta \Sigma }{\delta b^{a}}=\partial _{\mu }A_{\mu
}^{a}+gf^{abc}\lambda ^{b}c^{c}\;.  \label{gauge-fixing}
\end{equation}
}

\item  {The modified anti-ghost equation
\begin{equation}
\frac{\delta \Sigma }{\delta \bar{c}^{a}}+\partial _{\mu
}\frac{\delta \Sigma }{\delta \Omega _{\mu }^{a}}-gf^{abc}\lambda
^{b}\frac{\delta \Sigma }{\delta L^{c}}=-gf^{abc}\omega ^{b}c^{c}\;.
\label{anti-ghost}
\end{equation}
}

\item  {The modified ghost Ward identity
\begin{equation}
\label{ghosteq}\mathcal{G}^{a}(\Sigma )=\Delta
_{\mathrm{class.}}^{a}\;,
\end{equation}
with
\begin{eqnarray}
\mathcal{G}^{a}&=& \int d^{4}\!x\,\biggl(\frac{\delta }{\delta
c^{a}}+gf^{abc}\bar{c}^{b}\frac{\delta }{\delta b^{c}}+gf^{abc}\lambda ^{b}%
\frac{\delta }{\delta \omega ^{c}}\biggr)\;,\nonumber\\
\Delta _{\mathrm{class.}}^{a}&=&\int d^{4}\!x\,\Bigl[gf^{abc}%
\Bigl(\Omega _{\mu }^{b}A_{\mu }^{c}-L^{b}c^{c}-\omega ^{b}\bar{c}%
^{c}+(\beta -\rho )\lambda ^{b}\omega ^{c}-\lambda
^{b}b^{c}\Bigl)\Bigr] \nonumber\;.
\end{eqnarray}
}

\item  {The modified $\tau $-identity
\begin{equation}
\int d^{4}\!x\,\biggl(\frac{\delta \Sigma }{\delta \tau
}+c^{a}\frac{\delta
\Sigma }{\delta b^{a}}-2\lambda ^{a}\frac{\delta \Sigma }{\delta L^{a}}%
\biggr)=0 \;. \label{tau-equation}
\end{equation}
This identity expresses the on-shell BRST invariance of the operator
$A_{\mu }^a A_{\mu }^a$}.
\end{itemize}

\noindent We notice that every term on the r.h.s. of equations
(\ref{gauge-fixing})-(\ref{ghosteq}), being linear in the quantum
fields, represents a classical breaking \cite{Piguet:1995er}.

\noindent We are now prepared to write down the most general local
counterterm, $\Sigma _{\mathrm{ CT}}$, which is compatible with the
previous Ward identities and can be freely added to the original
action perturbatively. The perturbed action $\Sigma +\eta \Sigma
_{\mathrm{CT}}$ should obey the same Ward identities as the starting
action $\Sigma $ to the first order in the perturbation parameter
$\eta$. This corresponds to imposing the following constraints on
the counterterm $\Sigma_\mathrm{CT}$
\begin{itemize}
\item
\begin{equation}\label{constraint1}
\mathcal{B}_{\Sigma }\Sigma _{\mathrm{CT}}=0\;,
\end{equation}

\item
\begin{equation}\label{constraint2}
\frac{\delta \Sigma _{\mathrm{CT}}}{\delta b^{a}}%
=0 \;,
\end{equation}

\item
\begin{equation}\label{constraint3}
\frac{\delta \Sigma _{\mathrm{CT}}}{\delta \bar{%
c}^{a}}+\partial _{\mu }\frac{\delta \Sigma _{\mathrm{ CT}}}{\delta
\Omega _{\mu }^{a}}-gf^{abc}\lambda ^{b}\frac{\delta \Sigma _{\mathrm{ CT}%
}}{\delta L^{c}}=0\;,
\end{equation}

\item
\begin{equation}\label{constraint4}
\mathcal{G}^{a}(\Sigma _{\mathrm{CT}})=\int
d^{4}\!x\,\biggl(\frac{\delta \Sigma _{\mathrm{ CT}}}{\delta c^{a}}%
+gf^{abc}\lambda ^{b}\frac{\delta \Sigma _{\mathrm{ CT}}}{\delta
\omega ^{c}}\biggr)=0\;,
\end{equation}

\item
\begin{equation}\label{constraint5}
\int d^{4}\!x\,\biggl(\frac{\delta \Sigma _{%
\mathrm{ CT}}}{\delta \tau }-2\lambda ^{a}\frac{\delta \Sigma _{%
\mathrm{CT}}}{\delta L^{a}}\biggr)=0\;,
\end{equation}
\end{itemize}
\noindent where $\mathcal{B}_{\Sigma }$ is the nilpotent, $\mathcal{B}%
_{\Sigma }^{2}=0$, linearized Slavnov-Taylor operator, given by
\begin{equation}
\mathcal{B}_{\Sigma }=\int d^{4}\!x\,\biggl(\frac{\delta \Sigma
}{\delta
\Omega _{\mu }^{a}}\frac{\delta }{\delta A_{\mu }^{a}}+\frac{\delta \Sigma }{%
\delta A_{\mu }^{a}}\frac{\delta }{\delta \Omega _{\mu
}^{a}}+\frac{\delta
\Sigma }{\delta L^{a}}\frac{\delta }{\delta c^{a}}+\frac{\delta \Sigma }{%
\delta c^{a}}\frac{\delta }{\delta L^{a}}+b^{a}\frac{\delta }{\delta \bar{c}%
^{a}}+\omega ^{a}\frac{\delta }{\delta \lambda ^{a}}+J\frac{\delta
}{\delta \tau }\biggr).  \label{linearized}
\end{equation}
\noindent The most general local counterterm can be written as
\cite{Piguet:1995er}
\begin{equation}
\Sigma _{\mathrm{ CT}}=a_{0}\,S_{\mathrm{YM}}+\mathcal{B}_{\Sigma
}\Delta ^{-1},  \label{general-CT}
\end{equation}
\noindent where $\Delta ^{-1}$ is an integrated local polynomial of
ghost number $-1$ and dimension $4$, given by
\begin{eqnarray}
\Delta ^{-1} &=& \int d^{4}\!x \left[a_{1}\,\Omega _{\mu }^{a}A_{\mu
}^{a}+a_{2}\,L^{a}c^{a}+a_{3}\,A_{\mu
}^{a}\partial _{\mu }\bar{c}^{a}+a_{4}\,\frac{g}{2}f^{abc}\bar{c}^{a}\bar{c}%
^{b}c^{c}+a_{5}\,b^{a}\bar{c}^{a}+\frac{a_{6}}{2}\tau A_{\mu
}^{a}A_{\mu }^{a}\right.\nonumber\\  &+&\left.a_{7}\,\tau
\bar{c}^{a}c^{a}-\frac{a_{8}}{2}\zeta \tau J+a_{9}\,\gamma \lambda
^{a}\partial _{\mu }A_{\mu
}^{a}+a_{10}\,gf^{abc}\lambda ^{a}\bar{c}^{b}c^{c}+a_{11}\,\beta \frac{g}{2}%
f^{abc}\lambda ^{a}\lambda ^{b}c^{c}+a_{12}\,b^{a}\lambda
^{a}\right.\nonumber\\&+&\left.a_{13}\,\tau \lambda
^{a}c^{a}-\frac{a_{14}}{2}\rho \lambda ^{a}\omega
^{a}+a_{15}\,\omega ^{a}\bar{c}^{a}\right]\;.
\label{general-delta-1}
\end{eqnarray}
\noindent The constraints (\ref{constraint2})-(\ref{constraint5}),
imply that
\begin{eqnarray}
a_{1}&=&a_{3}=-a_{6}\nonumber\;,\\ a_{11}&=&\frac{\rho}{\beta}
a_{14}\nonumber\;, \\
a_{2}&=&a_{4}=a_{5}=a_{7}=a_{10}=a_{12}=a_{13}=a_{15}=0\;.
\label{a's}
\end{eqnarray}
\noindent Collecting these results, we come to the conclusion that
the most general counterterm compatible with the Ward identities
(\ref{constraint1})-(\ref{constraint5}) is eventually given by
\begin{eqnarray}
\Sigma _{\mathrm{ CT}} &=& \int d^{4}\!x\,%
\left\{\frac{a_{0}+2a_{1}}{2}\left[(\partial _{\mu }A_{\nu
}^{a})\partial _{\mu }A_{\nu }^{a}-(\partial _{\mu }A_{\nu
}^{a})\partial _{\nu }A_{\mu
}^{a}\right]+(a_{0}+3a_{1})\frac{g}{2}f^{abc}(\partial _{\mu }A_{\nu
}^{a})A_{\mu }^{b}A_{\nu }^{c}\right.\nonumber\\
&+&\left.(a_{0}+4a_{1})\frac{g^{2}}{4}f^{abc}f^{cde}A_{\mu
}^{a}A_{\nu }^{b}A_{\mu }^{d}A_{\nu }^{e}+a_{1}\,(\Omega _{\mu
}^{a}+\partial _{\mu }\bar{c}^{a})\partial _{\mu }c^{a}+\frac{a_{1}}{2}%
JA_{\mu }^{a}A_{\mu }^{a}\right.\nonumber\\
&+&\left.(a_{1}+a_{9})\gamma \omega ^{a}\partial _{\mu }A_{\mu
}^{a}+\rho a_{14}\,gf^{abc}\omega ^{a}\lambda ^{b}c^{c}+\rho a_{14}\,\frac{%
g^{2}}{4}f^{abc}f^{cde}\lambda ^{a}\lambda ^{b}c^{d}c^{e}\right.\nonumber\\
&-&\left.a_{14}\frac{\rho }{2}\omega ^{a}\omega ^{a}+\gamma
a_{9}\,\lambda ^{a}\partial ^{2}c^{a}+(a_{1}+a_{3})\gamma
gf^{abc}\lambda ^{a}\partial _{\mu }(A_{\mu
}^{b}c^{c})-a_{8}\,\frac{\zeta }{2}J^{2}\right\}\;. \label{final-CT}
\end{eqnarray}
\noindent Let us now check that this counterterm can be reabsorbed
in the original action (\ref{sigma}) by means of a multiplicative
renormalization of the available parameters, fields and sources,
according to
\begin{equation}
\Sigma (\Phi _{0},\phi _{0},\xi _{0})=\Sigma (\Phi ,\phi ,\xi )+\eta
\Sigma _{\mathrm{ CT}}(\Phi ,\phi ,\xi ) + O(\eta^2) \;,
\end{equation}
where
\begin{equation}
\Phi _{0}=Z_{\Phi }^{1/2}\Phi ,\qquad \phi _{0}=Z_{\phi }\phi
,\qquad \xi _{0}=Z_{\xi }\xi\;,   \label{renormalizations}
\end{equation}
with $\Phi =\{\Omega _{\mu }^{a},L^{a},\tau ,J,\lambda ^{a},\omega
^{a}\}$, $\phi =\{A_{\mu }^{a},b^{a},c^{a},\bar{c}^{a}\}$ and $\xi
=\{g,\zeta ,\rho ,\beta ,\gamma \}$. From equations
(\ref{final-CT})-(\ref {renormalizations}) one can check that
\begin{eqnarray}
Z_{g} &=&1-\eta \frac{a_{0}}{2}\nonumber\;, \\
Z_{A}^{1/2} &=& Z_{b}^{-1/2}=Z_{L}=Z_{\omega
}=1+\eta \Bigl(\frac{a_{0}}{2}+a_{1}\Bigr)\nonumber\;, \\
Z_{c}^{1/2} &=& Z_{\bar{c}}^{1/2}=Z_{\Omega
}=Z_{g}^{-1/2}Z_{A}^{-1/4}\nonumber\;, \\
Z_{J} &=&Z_{\tau }^{2}=Z_{g}Z_{A}^{-1/2}\nonumber\;, \\
Z_{\lambda } &=&Z_{g}^{-1/2}Z_{A}^{3/4}\nonumber\;, \\
Z_{\beta } &=& 1+\eta \Bigl(\frac{\rho
a_{14}}{\beta }-a_{0}-2a_{1}\Bigr)\nonumber\;, \\
Z_{\rho } &=& 1+\eta
(a_{14}-a_{0}-2a_{1})\nonumber\;, \\
Z_{\zeta } &=& 1+\eta
(a_{8}+2a_{0}+2a_{1})\nonumber\;, \\
Z_{\gamma } &=&1+\eta (a_{9}-a_{0}-a_{1})\;.
\end{eqnarray}\label{Z's}
allow for the desired multiplicative renormalization. We recall here
that the relations found for $Z_{J}$ and $Z_{\omega }$,
respectively, imply that the anomalous dimensions of the composite
operators $A_{\mu }^{2}$ and $gf^{abc}\bar{c}^{b}c^{c}$, are linear
combinations of that of $A_{\mu }^{a}$ and of the beta function
$\beta (g^{2})$ \cite{Dudal:2002pq,Dudal:2003dp}.

\sect{Effective potential for the condensates $\left\langle A_{\mu
}^{2}\right\rangle $ and $\left\langle \varepsilon ^{abc}%
\overline{c}^{b}c^{c}\right\rangle $.}
\subsection{Preliminaries.}
Let us now discuss how one can obtain the effective potential
describing the condensates $\left\langle A_{\mu
}^{2}\right\rangle $ and $\left\langle \varepsilon ^{abc}%
\overline{c}^{b}c^{c}\right\rangle $. Having proven the
renormalizability, we can set to zero all unnecessary external
sources, namely $\tau=0$, $\Omega_\mu^a=0$, $L^a=0$ and $\lambda=0$.
In this context, we would like to remark that we had to introduce
novel terms in $\lambda$ in the starting action,
eq.(\ref{supplement}). These terms, introduced for the algebraic
proof of the multiplicative renormalizability, are allowed by power
counting and by the symmetries of the model, and they do appear in
the most general counterterm, eq.(\ref{final-CT}). However, they are
not needed for the evaluation of the effective potential, so that
for our purposes $\lambda\equiv0$. Considering now the term
$\omega\partial A$ in eq.(\ref{supplement}), it is apparent that, in
the Landau gauge, $\p A=0$, such a term is absent. To make this
argument a little more formal, one could consider the generating
functional $\mathcal{W}(\omega,J)$ and perform the transformation of
the Lagrange multiplier $b^\prime=b+\gamma\omega$, with trivial
Jacobian.

\noindent We are still left with two free LCO parameters $\rho$ and
$\zeta$. As shown in
\cite{Verschelde:2001ia,Dudal:2003by,Dudal:2003dp}, these parameters
can be fixed by using the renormalization group equations. We would
like to notice that the explicit value of the counterterms $\propto
\omega^a$ are not influenced by the presence of $J$ and vice versa.
As such, the already determined values of $\rho$ and $\zeta$ remain
unchanged upon comparison with the cases $\omega^a\equiv0$
\cite{Verschelde:2001ia,Dudal:2003by} and $J\equiv0$
\cite{Dudal:2003dp}. More precisely, one still has
\begin{eqnarray}
\rho &=&\rho _{0}+\rho _{1}g^2+\ldots\;,  \nonumber \\
\zeta &=&\frac{\zeta _{0}}{g^{2}}+\zeta _{1}+\ldots\;,  \label{a2}
\end{eqnarray}
where
\begin{eqnarray}
\rho_0 &=&-\frac{6}{13}\;,\;\;\;\;\;\;\rho_1=-\frac{95}{312\pi^2} \;, \nonumber \\
\zeta_0 &=&\frac{27}{26}\;,\;\;\;\;\;\;\;\;\;\zeta_1= \frac{161}{52}\frac{3}{16\pi^2}\;, \label{a2bis}
\end{eqnarray}
in the case of $SU(2)$. The use of dimensional regularization with
$d=4-\varepsilon$ and of the $\MSbar$ renormalization scheme is
understood throughout this paper.

\noindent The relevant action is thus given by
\begin{eqnarray}
S&=&S_{\mathrm{YM}}+S_{\mathrm{GF}}+\int d^4x\left(gf^{abc}\omega^{a}\bar{c}%
^{b}c^{c}+\frac{1}{2}JA^{a}_{\mu}A^{a}_{\mu}-\frac{\rho}{2}\omega^a\omega^a-\frac{\zeta}{2}J^{2}\right)\;.
\label{Snodig}
\end{eqnarray}
As the sources $J$ and $\omega^a$ appear nonlinearly, the energy
interpretation might be spoiled. However, by exploiting a
Hubbard-Stratonovich transformation, the starting action
incorporating both gluon and ghost condensates can be rewritten as
\cite{Verschelde:2001ia,Dudal:2003dp,Dudal:2003by}
\begin{eqnarray}
S &=&S_{\mathrm{YM}}+S_{\mathrm{GF}}+\int d^{4}x\left( \frac{\phi ^{a}\phi ^{a}}{2g^{2}\rho }+
\frac{1}{\rho }\phi ^{a}\varepsilon ^{abc}\overline{c}^{b}c^{c}+\frac{g^{2}}{%
2\rho }\left( \varepsilon ^{abc}\overline{c}^{b}c^{c}\right)
^{2}\right)
\nonumber \\
&+&\int d^{4}x\left(\frac{\sigma ^{2}}{2g^{2}\zeta }+\frac{\sigma }{%
2g\zeta }A_{\mu }^{a}A_{\mu }^{a}+\frac{1}{8\zeta }\left( A_{\mu
}^{a}A_{\mu }^{a}\right) ^{2}-\omega^a\frac{\phi^a}{g}-J\frac{\sigma}{g}\right) \;,  \label{a1}
\end{eqnarray}
where the sources are now linearly coupled to the fields $\sigma$
and $\phi^a$, while the identifications
\begin{eqnarray}\label{ident}
  \left\langle\phi^a \right\rangle &=& -g^2\left\langle f^{abc}\oc^a c^b\right\rangle\;,\nonumber \\
  \left\langle \sigma \right\rangle &=&-\frac{g}{2}\left\langle
  A_\mu^2\right\rangle\;,
\end{eqnarray}
hold \cite{Verschelde:2001ia,Dudal:2003dp,Dudal:2003by}. The action
(\ref{a1}) will be multiplicatively renormalizable and will obey a
homogenous renormalization group equation.
\subsection{The one-loop effective potential.}
For the one-loop effective potential $V^{(1)}\left( \sigma ,\phi
\right)$ itself, we deduce
\begin{equation}
V^{(1)}\left( \sigma ,\phi \right) =V_{A^{2}}(\sigma )+V_{\mathrm{gh}}\left( \phi
\right) \;,  \label{a4}
\end{equation}
with \cite{Verschelde:2001ia,Dudal:2003by}
\begin{equation}
V_{A^{2}}(\sigma )=\frac{\sigma ^{2}}{2\zeta _{0}}\left( 1-\frac{%
\zeta _{1}}{\zeta _{0}}g^{2}\right) +\frac{3\left( N^{2}-1\right) }{%
64\pi ^{2}}\frac{g^{2}\sigma ^{2}}{\zeta _{0}^{2}}\left( \ln \frac{%
g\sigma }{\zeta _{0}\overline{\mu }^{2}}-\frac{5}{6}\right) \;,
\label{a5}
\end{equation}
while \cite{Dudal:2003dp}
\begin{equation}
V_{\mathrm{gh}}\left( \phi \right) =\frac{\phi ^{2}}{2g^{2}\rho _{0}}\left( 1-\frac{%
\rho _{1}}{\rho _{0}}g^{2}\right) +\frac{1}{32\pi ^{2}}\frac{\phi ^{2}}{\rho
_{0}^{2}}\left( \ln \frac{\phi ^{2}}{\rho _{0}^{2}\overline{\mu }^{4}}%
-3\right) \;,  \label{a6}
\end{equation}
where $\phi =\phi ^{3}$, $\phi ^{a}=\phi \delta ^{a3}$. This amounts
to choosing the vacuum configuration along the 3-direction in color
space. For the rest of the paper, it is understood that $N=2$.

\noindent The minimum configuration, describing the vacuum, is
retrieved by solving
\begin{equation}
\frac{\partial V^{(1)}\left( \sigma ,\phi \right) }{\partial \sigma }=\frac{%
\partial V^{(1)}\left( \sigma ,\phi \right) }{\partial \phi }=0\;,
\label{a7}
\end{equation}
or
\begin{eqnarray}
\frac{1}{\zeta _{0}}\left( 1-\frac{\zeta _{1}}{\zeta _{0}}%
g^{2}\right) +2\frac{9}{64\pi ^{2}}\frac{g^{2}}{%
\zeta _{0}^{2}}\left( \log \frac{g\sigma _{*}}{\zeta _{0}\overline{%
\mu }^{2}}-\frac{5}{6}\right) +\frac{9 }{64\pi ^{2}}%
\frac{g^{2}}{\zeta _{0}^{2}} &=&0\;,  \nonumber \\
\frac{1}{g^{2}\rho _{0}}\left( 1-\frac{\rho _{1}}{\rho _{0}}g^{2}\right) +2%
\frac{1}{32\pi ^{2}}\frac{1}{\rho _{0}^{2}}\left( \log \frac{\phi _{*}^{2}}{%
\rho _{0}^{2}\overline{\mu }^{4}}-3\right) +2\frac{1}{32\pi ^{2}}\frac{1}{%
\rho _{0}^{2}} &=&0\;,  \label{a8}
\end{eqnarray}
where $\left( \sigma _{*},\phi _{*}\right)$ denote the nontrivial
solution. For the vacuum energy, we obtain
\begin{equation}
\Evac =V^{(1)}\left( \sigma
_{*},\phi _{*}\right) =-\frac{9}{128\pi ^{2}}\frac{%
g^{2}}{\zeta _{0}^{2}}\sigma _{*}^{2}-\frac{1}{32\pi ^{2}}\frac{\phi
_{*}^{2}}{\rho _{0}^{2}}\;.  \label{a9}
\end{equation}
One sees thus that nonvanishing condensates will be dynamically
favoured at one-loop as they both lower the vacuum energy.

\noindent For further investigation, it is useful to introduce the
variables
\begin{eqnarray}
m^{2} &=&\frac{g\sigma }{\zeta _{0}}\;,  \nonumber \\
\omega &=&\frac{\phi }{\left|\rho _{0}\right|}\;.  \label{a15b}
\end{eqnarray}
hence
\begin{equation}
V_{A^{2}}(m^2 )=\zeta_0\frac{m ^{4}}{2g^2}\left( 1-\frac{%
\zeta _{1}}{\zeta _{0}}g^{2}\right) +\frac{3\left( N^{2}-1\right) }{%
64\pi ^{2}}m^4\left( \ln \frac{m^2}{\overline{\mu
}^{2}}-\frac{5}{6}\right) \;, \label{ll1}
\end{equation}
while
\begin{equation}
V_{\mathrm{gh}}\left( \omega \right) =-\rho_0\frac{\omega ^{2}}{2g^{2}}\left( 1-\frac{%
\rho _{1}}{\rho _{0}}g^{2}\right) +\frac{\omega^2}{32\pi
^{2}}\left( \ln \frac{\omega ^{2}}{\overline{\mu }^{4}}%
-3\right) \;,  \label{ll2}
\end{equation}
\noindent Let us now try to get an estimate of the vacuum state of
the theory. In comparison with the case where only the gluon or
ghost condensation is considered, we have now an additional
complication, due to the presence of two mass scales. Usually, when
only a single scale is present, one chooses the renormalization
scale $\omu^2$ in such a way that potentially large logarithms
vanish in the gap equation. In the present case, two different
logarithms show up. In order to keep some control on the expansion,
we shall use the RG invariance to explicitly sum the leading
logarithms (LL) in the effective potential. To this end, we notice
that the potential can be rewritten as
\begin{eqnarray}\label{ll3}
    V&=&\zeta_0\frac{m^4}{g^2}\sum_{n=0}^{\infty}a_n
    \left(g^2\ln\frac{m^2}{\omu^2}\right)^n+m^4\left(-\frac{\zeta_1}{2}-\frac{5}{6}\frac{3(N^2-1)}{64\pi^2}\right)\;,\nonumber\\
    &+&\rho_0\frac{\omega^2}{g^2}\sum_{n=0}^{\infty}b_n
    \left(g^2\ln\frac{\omega^2}{\omu^4}\right)^n+\omega^2\left(\frac{\rho_1}{2}-\frac{3}{32\pi^2}\right)\;,
\end{eqnarray}
for a LL-expansion\footnote{A term like
$m^4\left(g^2\ln\frac{\omega^2}{\omu^4}\right)$ or
$\omega^2\left(g^2\ln\frac{m^2}{\omu^2}\right)$ shall not occur, as
there are no infrared divergences for $m^2=0$, $\omega\neq0$ or
$m^2\neq0$ or $\omega=0$.}, where $a_0=-b_0=\frac{1}{2}$. For the
time being, we shall only consider the part in $m^2$. The analysis
for the part in $\omega$ is completely analogous and independent
from the $m^2$-part. \noindent We set
\begin{eqnarray}\label{ll4}
  \omu\frac{\p}{\p\omu}g^2 &=& \beta(g^2)=-2\sum_{n=0}^{\infty}\beta_n (g^2)^{n+2}\;,\nonumber \\
  \omu\frac{\p}{\p\omu}\ln m^2 &=& \gamma_2(g^2)=\sum_{n=0}^{\infty}\gamma_n (g^2)^{n+1}\;.
\end{eqnarray}
Since we know that the effective potential should be RG invariant,
we find that
\begin{eqnarray}\label{ll5}
    \omu\frac{\p}{\p\omu}V&=&0\nonumber\\&\Downarrow&\nonumber\\(\gamma_0+\beta_0)\sum_{n=0}^{\infty}a_n
    u^n-(\beta_0u+1)\sum_{n=0}^{\infty}(n+1)a_{n+1}u^n&=&0+\textrm{next-to-leading order}\;.\nonumber\\
\end{eqnarray}
We have defined
\begin{equation}\label{ll5bis}
    u=g^2\ln\frac{m^2}{\omu^2}\;.
\end{equation}
Setting
\begin{equation}\label{ll6}
    F(u)=\sum_{n=0}^{\infty}a_{n}u^n\;,
\end{equation}
then eq.(\ref{ll5}) translates into a differential equation
\begin{equation}
    (\gamma_0+\beta_0)F(u)-(\beta_0u+1)F^\prime(u)=0\;,
\end{equation}
which can be solved to
\begin{equation}\label{ll7}
    F(u)=\frac{1}{2}\left(1+\beta_0u\right)^{\frac{\gamma_0+\beta_0}{\beta_0}}\;,
\end{equation}
as we have the initial condition $F(0)=\frac{1}{2}$. Using this
result, we can write
\begin{eqnarray}
V_{A^{2}}(m^2 )&=&\zeta_0\frac{m
^{4}(\omu)}{2g^2(\omu)}\frac{\left(1+\beta_0u\right)^{\frac{\gamma_0}{\beta_0}}}{\left(1+\beta_0u\right)^{-1}}+m^4(\omu)\left(-\frac{\zeta_1}{2}-\frac{5}{6}\frac{3(N^2-1)}{64\pi^2}\right)+\cdots
\nonumber\\
&=&\zeta_0\frac{m
^{4}(m)}{2g^2(m)}+m^4(m)\left(-\frac{\zeta_1}{2}-\frac{5}{6}\frac{3(N^2-1)}{64\pi^2}\right)+\cdots\;,
\label{ll8}
\end{eqnarray}
as the running quantities at scale $\omu$ get replaced by their
counterparts at scale $m$.

\noindent The same could be done for the ghost condensation part, so
that we can write for the LL summed effective potential
\begin{eqnarray}\label{ll9}
    V&=&\frac{\zeta_0}{2}\frac{\om^4}{\og^2}+\om^4\left(-\frac{\zeta_1}{2}-\frac{5}{6}\frac{3(N^2-1)}{64\pi^2}\right)-\frac{\rho_0}{2}\frac{\widetilde{\omega}^2}{\widetilde{g}^2}+\widetilde{\omega}^2\left(\frac{\rho_1}{2}-\frac{3}{32\pi^2}\right)\;.
\end{eqnarray}
where it is understood that the barred quantities like $\og^2$ are
considered at scale $\omu^2=m^2$, and the tilded quantities like
$\widetilde{g}^2$ at scale $\omu^2=\omega$.

\noindent For further usage, let us first quote the explicit values
of the anomalous dimensions of $g^2$, $m^2$ and $\omega$. Using the
definitions (\ref{ll4}) one shall find in e.g. \cite{Dudal:2005na}
that
\begin{eqnarray}\label{ll10}
\beta_0=\frac{11}{3}\frac{N}{16\pi^2}\;,\nonumber\\
\gamma_0=-\frac{3}{2}\frac{N}{16\pi^2}\;.
\end{eqnarray}
Defining
\begin{eqnarray}\label{ll4bis}
  \omu\frac{\p}{\p\omu}\ln\omega &=& \kappa(g^2)=\sum_{n=0}^{\infty}
  \kappa_n(g^2)^{n+1}\;,
\end{eqnarray}
one can infer from \cite{Dudal:2003by} and the definitions
(\ref{ident})-(\ref{a15b}) that
\begin{equation}\label{ll11}
    \kappa(g^2)=\frac{1}{2}\frac{\beta(g^2)}{g^2}+\gamma_A(g^2)\;,
\end{equation}
where $\gamma_A(g^2)$ is the anomalous dimension of the gluon field
in the Landau gauge as defined in \cite{Dudal:2003by}. Hence,
\begin{equation}\label{ll12}
    \kappa_0=-\frac{35}{6}\frac{N}{16\pi^2}\;.
\end{equation}
We are now ready to determine the minimum configuration. The gap
equations we intend to solve are given by
\begin{eqnarray}\label{ll13}
  \frac{\p V}{\p \om^2} = 0&\Rightarrow& \frac{\zeta_0}{\og^2}+\zeta_0(\beta_0+\gamma_0)-2\left(\frac{\zeta_1}{2}+\frac{5(N^2-1)}{128\pi^2}\right)=0\;,\nonumber\\
  \frac{\p V}{\p \widetilde{\omega}} =
  0&\Rightarrow&-\frac{\rho_0}{\widetilde{g}^2}-\rho_0(\beta_0+\kappa_0)+2\left(\frac{\rho_1}{2}-\frac{3}{32\pi^2}\right)=0\;.
\end{eqnarray}
or
\begin{eqnarray}\label{ll14}
\left.\frac{\og^2N}{16\pi^2}\right|_{N=2}&=&\frac{9}{37}\approx0.243\;,\nonumber\\
\left.\frac{\widetilde{g}^2N}{16\pi^2}\right|_{N=2}&=&\frac{36}{385}\approx0.094\;.
\end{eqnarray}
Using the one-loop $\MSbar$ expression
\begin{equation}\label{ll15}
    g^2(\omu)=\frac{1}{\beta_0\ln\frac{\omu^2}{\lms^2}}\;,
\end{equation}
one extracts from the values (\ref{ll14}) the estimates
\begin{eqnarray}\label{ll16}
\overline{m}^2=e^{\frac{37}{33}}\lms^2&\approx&\nonumber3.07\lms^2\;,\\
\widetilde{\omega}=e^{\frac{35}{12}}\lms^2&\approx&18.48\lms^2\;,
\end{eqnarray}
while for the vacuum energy, eq.(\ref{a9}), we obtain
\begin{equation}\label{ll17}
    \Evac\approx-1.15\lms^4\;.
\end{equation}
As the RG improved coupling constants of eq.(\ref{ll14}) are
relatively small, the performed expansion should have some
trustworthiness. Evidently, explicit knowledge of the higher order
contributions would be necessary to reach better conclusions about
the reliability of the presented values, but this is beyond the
scope of this article.

\sect{A study of the one-loop ghost contribution to the vacuum
polarization.} We shall now start to investigate the consequences
stemming from a nonvanishing condensate $\left\langle
\varepsilon^{3bc}\oc^b c^c\right\rangle$. Let us begin by making a
detailed study of the one-loop ghost contribution to the vacuum
polarization. Before starting with the explicit computations, it is
worth giving a look at the Ward identity obeyed by the one-loop
vacuum polarization stemming from the Slavnov-Taylor identity
describing the theory in the condensed vacuum. To focus on the role
of the ghost
condensate, we switch off, for the time being, the gluon condensate $%
\left\langle A_{\mu }^{2}\right\rangle $. Its inclusion can be done
straightforwardly. Let us start thus with the action
\begin{equation}
S=S_{\mathrm{YM}}+S_{\mathrm{GF}}+\int d^{4}x\left( \frac{\phi ^{a}\phi ^{a}}{2g^{2}\rho }+
\frac{1}{\rho }\phi ^{a}\varepsilon ^{abc}\overline{c}^{b}c^{c}+\frac{g^{2}}{%
2\rho }\left( \varepsilon ^{abc}\overline{c}^{b}c^{c}\right)
^{2}\right) \;, \label{b1}
\end{equation}
where
\begin{eqnarray}
\phi ^{a}(x) &=&\delta ^{a3}\phi _{*}+\widetilde{\phi }^{a}(x)\;,  \nonumber
\\
\left\langle \widetilde{\phi }^{a}(x)\right\rangle &=&0\;.  \label{b2}
\end{eqnarray}
Requiring that
\begin{equation}
s\phi ^{a}=s\widetilde{\phi }^{a}=-g^{2}s(\varepsilon ^{abc}\overline{c}%
^{b}c^{c})\;,  \label{b4}
\end{equation}
will assure that $S$ is left invariant by the following nilpotent BRST
transformations
\begin{eqnarray}
sA_{\mu }^{a} &=&-D_{\mu }^{ab}c^{b}\;,  \nonumber \\
sc^{a} &=&\frac{g}{2}\varepsilon ^{abc}c^{b}c^{c}\;,  \nonumber \\
s\widetilde{\phi }^{a} &=&-g^{2}\left( \varepsilon ^{abc}b^{b}c^{c}+\frac{g}{%
2}\varepsilon ^{abc}\overline{c}^{b}\varepsilon
^{cmn}c^{m}c^{n}\right) \;,
\nonumber \\
s\phi _{*} &=&0\;,  \nonumber \\
s\overline{c}^{a} &=&b^{a}\;,  \nonumber \\
sb^{a} &=&0\;,  \label{b5}
\end{eqnarray}
thus
\begin{equation}
sS=0\;.  \label{b6}
\end{equation}
In order to obtain the Slavnov-Taylor identity, we introduce external
sources $\Omega _{\mu }^{a}$, $L^{a}$, $F^{a}$ coupled to the nonlinear
variations of the fields.
\begin{equation}
S_{\mathrm{ext}}=\int d^{4}x\left( -\Omega _{\mu }^{a}D_{\mu }^{ab}c^{b}+\frac{g}{2}%
L^{a}\varepsilon ^{abc}c^{b}c^{c}+F^{a} s\widetilde{\phi
}^{a}\right) \;,  \label{b7}
\end{equation}
The complete action
\begin{equation}
\Sigma =S+S_{\mathrm{ext}}\;,  \label{b8}
\end{equation}
obeys the Slavnov-Taylor identity
\begin{equation}
\mathcal{S}(\Sigma )=0\;,  \label{b9}
\end{equation}
where
\begin{equation}
\mathcal{S}(\Sigma )=\int d^{4}x\left( \frac{\delta \Sigma }{\delta \Omega
_{\mu }^{a}}\frac{\delta \Sigma }{\delta A_{\mu }^{a}}+\frac{\delta \Sigma }{%
\delta L^{a}}\frac{\delta \Sigma }{\delta c^{a}}+\frac{\delta \Sigma }{%
\delta F^{a}}\frac{\delta \Sigma }{\delta \widetilde{\phi }^{a}}+b^{a}\frac{%
\delta \Sigma }{\delta \overline{c}^{a}}\right) \;.  \label{b10}
\end{equation}
Due to the absence of anomalies and to the stability of the theory,
the Slavnov-Taylor identity ({\ref{b9}}) holds at the quantum level,
i.e.
\begin{equation}
\mathcal{S}(\Gamma )=\int d^{4}x\left( \frac{\delta \Gamma }{\delta \Omega
_{\mu }^{a}}\frac{\delta \Gamma }{\delta A_{\mu }^{a}}+\frac{\delta \Gamma }{%
\delta L^{a}}\frac{\delta \Gamma }{\delta c^{a}}+\frac{\delta \Gamma }{%
\delta F^{a}}\frac{\delta \Gamma }{\delta \widetilde{\phi }^{a}}+b^{a}\frac{%
\delta \Gamma }{\delta \overline{c}^{a}}\right) =0\;,  \label{b11}
\end{equation}
where $\Gamma $
\begin{equation}
\Gamma =\Sigma +\hbar \Gamma ^{1}+\ldots\;,  \label{b12}
\end{equation}
is the generator of the $1PI$ Green functions.

\subsection{Ward identity for the vacuum polarization.}
Let us now derive the Ward identity for the vacuum polarization at
one-loop
level following from the Slavnov-Taylor identity $\left( {\ref{b11}}\right) $%
. At one-loop level, one has
\begin{equation}
\Gamma =\Sigma +\hbar \Gamma ^{1}\;,  \label{b13}
\end{equation}
so that the Slavnov-Taylor identity becomes
\begin{eqnarray}
\int d^{4}x\left( \frac{\delta \Gamma ^{1}}{\delta \Omega _{\mu }^{a}}\frac{%
\delta \Sigma }{\delta A_{\mu }^{a}}+\frac{\delta \Sigma }{\delta \Omega
_{\mu }^{a}}\frac{\delta \Gamma ^{1}}{\delta A_{\mu }^{a}}+\frac{\delta
\Gamma ^{1}}{\delta L^{a}}\frac{\delta \Sigma }{\delta c^{a}}+\frac{\delta
\Sigma }{\delta L^{a}}\frac{\delta \Gamma ^{1}}{\delta c^{a}}+\frac{\delta
\Gamma ^{1}}{\delta F^{a}}\frac{\delta \Sigma }{\delta \widetilde{\phi }^{a}}%
+\frac{\delta \Sigma }{\delta F^{a}}\frac{\delta \Gamma ^{1}}{\delta
\widetilde{\phi }^{a}}+b^{a}\frac{\delta \Gamma ^{1}}{\delta \overline{c}^{a}%
}\right) =0\;.\nonumber\\\label{stid}
\end{eqnarray}
From
\begin{eqnarray}
\frac{\delta \Gamma ^{1}}{\delta \Omega _{\mu }^{a}} &=&\left[ -\left(
D_{\mu }^{ab}c^{b}\right) \cdot \Gamma \right] ^{1}\;,  \nonumber \\
\frac{\delta \Gamma ^{1}}{\delta L^{a}} &=&\left[ \left( \frac{g}{2}%
\varepsilon ^{abc}c^{b}c^{c}\;\right) \cdot \Gamma \right] ^{1}\;,
\nonumber
\\
\frac{\delta \Gamma ^{1}}{\delta F^{a}} &=&\left[ -g^{2}\left(
\varepsilon ^{abc}b^{b}c^{c}+\frac{g}{2}\varepsilon
^{abc}\overline{c}^{b}\varepsilon ^{cmn}c^{m}c^{n}\right) \;\cdot
\Gamma \right] ^{1}\;,  \label{b14}
\end{eqnarray}
where $\left[ \mathcal{O}\cdot \Gamma \right] $ denotes the generator of the $1PI$ Green
functions with the insertion of the composite operator $\mathcal{O}$, it follows that
\begin{eqnarray}
0 &=&\int d^{4}x\left( \left[ -\left( D_{\mu }^{ab}c^{b}\right)
\cdot \Gamma \right] ^{1}\frac{\delta \Sigma }{\delta A_{\mu
}^{a}}-\left( D_{\mu }^{ab}c^{b}\right) \frac{\delta \Gamma
^{1}}{\delta A_{\mu }^{a}}+\left[ \left( \frac{g}{2}\varepsilon
^{abc}c^{b}c^{c}\;\right) \cdot \Gamma \right]
^{1}\frac{\delta \Sigma }{\delta c^{a}}\right.  \nonumber \\
&+&\left( \frac{g}{2}\varepsilon ^{abc}c^{b}c^{c}\;\right) \frac{%
\delta \Gamma ^{1}}{\delta c^{a}}+\left[ -g^{2}\left( \varepsilon
^{abc}b^{b}c^{c}+\frac{g}{2}\varepsilon
^{abc}\overline{c}^{b}\varepsilon
^{cmn}c^{m}c^{n}\right) \;\cdot \Gamma \right] ^{1}\frac{\delta \Sigma }{%
\delta \widetilde{\phi }^{a}}  \nonumber \\
&-&\left. g^{2}\left( \varepsilon ^{abc}b^{b}c^{c}+\frac{g}{2}%
\varepsilon ^{abc}\overline{c}^{b}\varepsilon ^{cmn}c^{m}c^{n}\right) \frac{%
\delta \Gamma ^{1}}{\delta \widetilde{\phi }^{a}}+b^{a}\frac{\delta \Gamma
^{1}}{\delta \overline{c}^{a}}\right)\;.  \label{b15}
\end{eqnarray}
Acting on both sides of eq.({\ref{b15}) with the test
operator
\begin{equation}
\frac{\delta ^{2}}{\delta c^{a}(x)\delta A_{\nu }^{b}(y)}\;,  \label{b16}
\end{equation}
and setting all fields and sources equal to zero, $A_{\mu }^{a}=\Omega _{\mu
}^{a}=c^{a}=L^{a}=\widetilde{\phi }^{a}=F^{a}=b^{a}=\overline{c}^{a}=0$, one
obtains the Ward identity for the vacuum polarization in the condensed ghost
vacuum
\begin{equation}
\partial _{\mu }^{x}\frac{\delta ^{2}\Gamma ^{1}}{\delta A_{\mu
}^{a}(x)\delta A_{v}^{b}(y)}=\frac{\phi _{*}}{\rho _{0}}\left(
\frac{\delta ^{2}\left[ \int d^{4}z\left( \varepsilon
^{3np}b^{n}c^{p}\right)
_{z}\;\cdot \Gamma \right] ^{1}}{\delta c^{a}(x)\delta A_{\nu }^{b}(y)}%
\right) \;.  \label{b17}
\end{equation}
The right hand side of equation ({\ref{b17}) is the one-loop $1PI$
Green function with the insertion of the integrated composite
operator $\int d^{4}z\left( \varepsilon ^{3np}b^{n}c^{p}\right)
_{z}$, with one gluon and one ghost as (amputated) external legs.
This Green function, shown in Figure 1, is nonvanishing.
\begin{figure}[ht]
\begin{center}
        \scalebox{1}{\includegraphics{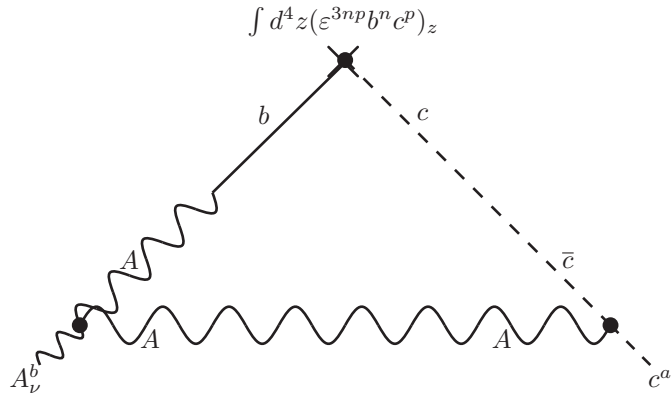}}
        \caption{The Green function appearing in the r.h.s. of eq.(\ref{b17}).}
    \end{center}
\end{figure}

\noindent For instance, for the Fourier transform of the component
$a=b=3$, one finds
\begin{equation}
\left( \frac{\delta ^{2}\left[ \int d^{4}x_{1}\left( \varepsilon
^{3np}b^{n}c^{p}\right) _{x_{1}}\;\cdot \Gamma \right] ^{1}}{\delta
c^{3}\delta A_{\nu }^{3}}\right) =\mathcal{M}_{\nu }^{33}(p,\omega
)\;, \label{b17b}
\end{equation}
with
\begin{equation}
\mathcal{M}_{\nu }^{33}(p,\omega )=-2g^{2}\int \frac{d^{d}k}{\left( 2\pi
\right) ^{d}}\frac{\left( k^{2}\delta _{\nu \rho }+p_{\nu }p_{\rho }-2k_{\nu
}p_{\rho }\right) p_{\tau }}{\left( k^{4}+\omega ^{2}\right) \left( \left(
p-k\right) ^{2}+m^{2}\right) }\left( \delta _{\rho \tau }-\frac{\left(
p-k\right) _{\rho }\left( p-k\right) _{\tau }}{\left( p-k\right) ^{2}}%
\right) \;,  \label{b18}
\end{equation}
where use has been made of the mixed $b-A$ propagator
\begin{equation}
\left\langle b^{a}(k)A_{\mu }^{b}(-k)\right\rangle =\frac{k_{\mu }}{k^{2}}%
\;.\label{mixedprop}
\end{equation}
From the identity
\begin{equation}
\frac{k^{2}}{k^{4}+\omega ^{2}}=\frac{1}{k^{2}}-\frac{\omega ^{2}}{\left(
k^{4}+\omega ^{2}\right) k^{2}}\;,  \label{b19}
\end{equation}
it follows that
\begin{equation}
\mathcal{M}_{\nu }^{33}(p,\omega )=\mathcal{M}_{\nu }^{33}(p,0)+\omega ^{2}%
\widetilde{\mathcal{M}}_{\nu }^{33}(p,\omega )\;,  \label{b20}
\end{equation}
where $\mathcal{M}_{\nu }^{33}(p,0)$ is logarithmic divergent, while $%
\widetilde{\mathcal{M}}_{\nu }^{33}(p,\omega )$ is ultraviolet
finite. One should certainly notice that the right hand side of
eq.({\ref{b17}}) is proportional to the ghost condensation $\phi
_{*}$.

\noindent In summary, the Ward identity $\left( {\ref{b17}}\right) $
shows at a formal level that the one-loop vacuum polarization in the
condensed ghost vacuum is not transverse. This will be explicitly
checked in section 6 at one-loop order. We recognize that, in the
absence of the ghost condensation, the well known result of the
transversality of the vacuum polarization is recovered.

\noindent We observe that, due to Lorentz invariance, one may write
\begin{equation}
\frac{\phi _{*}}{\rho _{0}}\mathcal{M}_{\nu }^{33}(p,\omega
)=a^{33}(p,\omega )p_{\nu }\;,  \label{b21}
\end{equation}
where $a^{33}(p,\omega )$ is a suitable scalar quantity. The Ward identity $%
\left( {\ref{b17}}\right) $ becomes thus
\begin{equation}
p_{\mu }\Pi _{\mu \nu }^{33}(p,\omega )=a^{33}(p,\omega )p_{\nu }\;,
\label{b22}
\end{equation}
where $\Pi _{\mu \nu }^{33}(p,\omega )$ stands for the vacuum
polarization. Equation $\left( {\ref{b22}}\right) $ can be recast
into the form
\begin{equation}
p_{\mu }\left( \Pi _{\mu \nu }^{33}(p,\omega )-a^{33}(p,\omega )\delta _{\mu
\nu }\right) =0\;,  \label{b23}
\end{equation}
which is suitable for analyzing the location of the pole of the
complete gluon propagator $\mathcal{G}_{\mu \nu }(p)$ in the
condensed vacuum.

\sect{Ward identity for the gluon propagator.} Having discussed the
breakdown of the transversality of the vacuum polarization due to
the ghost condensation, one might wonder what happens with the gluon
propagator itself in the ghost condensed vacuum.
\subsection{Consequences of the ghost condensation on the pole of the propagator.}
In order to have a better understanding of the Ward identity $\left(
{\ref {b17}}\right) $, let us discuss the resulting modification on
the location of the pole of the gluon propagator\ $\mathcal{G}_{\mu
\nu }(p)$. From eq.(\ref{b23}), it follows that
\begin{equation}
\Pi _{\mu \nu }^{33}(p,\omega )=\left( \delta _{\mu \nu }-\frac{p_{\mu
}p_{\nu }}{p^{2}}\right) \Pi ^{33}(p,\omega )+a^{33}(p,\omega )\delta _{\mu
\nu }\;.  \label{w4}
\end{equation}
Therefore, for the complete one-loop propagator $\mathcal{G}_{\mu
\nu }^{33}(p)$ one has
\begin{equation}
\mathcal{G}_{\mu \nu }^{33}(p)=\frac{P_{\mu \nu }(p)}{p^{2}+m^{2}}-\frac{%
P_{\mu \rho }(p)}{p^{2}+m^{2}}\Pi _{\rho \sigma }^{33}(p,\omega )\frac{%
P_{\sigma \nu }(p)}{p^{2}+m^{2}}\;,  \label{w5}
\end{equation}
where
\begin{equation}
P_{\mu \nu }(p)= \delta _{\mu \nu }-\frac{p_{\mu
}p_{\nu}}{p^{2}}\;,\label{transproj}
\end{equation}
is the transversal projector. Thus
\begin{eqnarray}
\mathcal{G}_{\mu \nu }^{33}(p) &=&\frac{P_{\mu \nu }(p)}{p^{2}+m^{2}}-\frac{%
P_{\mu \rho }(p)}{p^{2}+m^{2}}P_{\rho \sigma }(p)\Pi ^{33}(p,\omega )\frac{%
P_{\sigma \nu }(p)}{p^{2}+m^{2}}-\frac{P_{\mu \rho }(p)}{p^{2}+m^{2}}%
a^{33}(p,\omega )\delta _{\rho \sigma }\frac{P_{\sigma \nu }(p)}{p^{2}+m^{2}}
\nonumber \\
&=&\frac{P_{\mu \nu }(p)}{p^{2}+m^{2}}-\frac{P_{\mu \nu }(p)}{p^{2}+m^{2}}%
\frac{\Pi ^{33}(p,\omega )}{p^{2}+m^{2}}-\frac{P_{\mu \nu }(p)}{p^{2}+m^{2}}%
\frac{a^{33}(p,\omega )}{p^{2}+m^{2}}  \nonumber \\
&=&\frac{P_{\mu \nu }(p)}{p^{2}+m^{2}}\left( 1-\frac{\left( \Pi
^{33}(p,\omega )+a^{33}(p,\omega )\right) }{p^{2}+m^{2}}\right) \simeq \frac{%
P_{\mu \nu }(p)}{p^{2}+m^{2}}\frac{1}{1+\frac{\left( \Pi ^{33}(p,\omega
)+a^{33}(p,\omega )\right) }{p^{2}+m^{2}}}\;.  \label{w6}
\end{eqnarray}
Finally
\begin{equation}
\mathcal{G}_{\mu \nu }^{33}(p)\simeq \frac{P_{\mu \nu }(p)}{p^{2}+m^{2}+\Pi
^{33}(p,\omega )+a^{33}(p,\omega )}\;,  \label{w7}
\end{equation}
from which one sees that the location of the pole is indeed affected
by the presence of the quantity $a^{33}(p,\omega )$.

\noindent Analogously, for the
components $\left( \alpha ,\beta \right) $ of the gluon propagator, one
shall find that
\begin{equation}
\mathcal{G}_{\mu \nu }^{\alpha \beta }(p)\simeq \delta ^{\alpha \beta }\frac{%
P_{\mu \nu }(p)}{p^{2}+m^{2}+\Pi (p,\omega )+a(p,\omega )}\;,  \label{w8}
\end{equation}
with
\begin{equation}
\Pi _{\mu \nu }^{\alpha \beta }(p,\omega )=\delta ^{\alpha \beta }\left(
\left( \delta _{\mu \nu }-\frac{p_{\mu }p_{\nu }}{p^{2}}\right) \Pi
(p,\omega )+a(p,\omega )\delta _{\mu \nu }\right) \;.  \label{w9}
\end{equation}
Equations ({\ref{w7}}), ({\ref{w8}}) express the physical meaning of
the Ward identity ({\ref{b17}}), namely, due to the violation of the
transversality of the vacuum polarization in the condensed vacuum,
the location of the pole of the propagator is affected by the
quantity $\mathcal{M}_{\nu }(p,\omega )$ appearing in the right hand
side of eq.({\ref{b17}). We also notice that the gluon propagator,
being proportional to the projector $P_{\mu\nu}(p)$,
({\ref{transproj}}), remains transverse. This will be formally
proven in the following subsection.

\subsection{Transversality of the gluon propagator.}
As the gluon propagator is the connected two-point function, we
should consider the generator $Z^c$ of connected Green functions,
obtained from the $1PI$ quantum action $\Gamma$ by means of a
Legendre transformation. Therefore, we introduce sources $I_b^a$,
$J_\mu^a$ and $K_c^a$, respectively for the fields $b^a$, $A_\mu^a$
and $c^a$. The effective action (\ref{b12}) obeys the Ward identity
\begin{equation}\label{gp1}
    \frac{\delta\Gamma}{\delta b^a}=\p_\mu A_\mu^a+g^2
    \varepsilon^{ade}F^d c^e\;.
\end{equation}
At the level of $Z^c$, this identity is translated into
\begin{equation}\label{gp2}
    I_b^a=\p_\mu\frac{\delta Z^c}{\delta
    J_\mu^a}+g^2\varepsilon^{ade}F^d\frac{\delta Z^c}{\delta K_c^e}\;,
\end{equation}
from which one deduces
\begin{equation}\label{gp3}
    0=\p_\mu^x\left.\frac{\delta^2 Z^c}{\delta J_\mu^a(x)
    \delta
    J_\mu^b(y)}\right|_{F,I,J,K=0}=\p_\mu^x\mathcal{G}_{\mu\nu}^{ab}(x,y)\;,
\end{equation}
meaning that the gluon propagator does remain transverse in the
ghost condensed vacuum.

\sect{One-loop evaluation of the ghost contribution to the vacuum
polarization.} In this section, we shall discuss the one-loop
contribution to the vacuum polarization coming from diagrams that
contain internal ghost lines, denoted by $\left( \Pi _{\mu \nu
}^{ab}(p,\omega )\right) _{\mathrm{gh}}$.

\noindent In order to evaluate the one-loop ghost contribution to
the vacuum polarization, let us first remind the form of the tree
level gluon and ghost propagators in the nontrivial vacuum, given by
$\left\langle A_{\mu
}^2\right\rangle \neq 0$ and $\left\langle \varepsilon ^{abc}%
\overline{c}^{b}c^{c}\right\rangle \neq 0$. From
\cite{Verschelde:2001ia,Dudal:2003by}, the gluon propagator reads
\begin{equation}
\left\langle A_{\mu }^{a}(k)A_{\nu }^{b}(-k)\right\rangle =\delta ^{ab}\frac{%
1}{k^{2}+m^{2}}\left( \delta _{\mu \nu }-\frac{k_{\mu }k_{\nu }}{k^{2}}%
\right) \;,\;\;\;\;\;\;\;a,b=1,2,3\;.  \label{c1}
\end{equation}
Here we see the meaning of the parameter $m^2$, defined in
eq.(\ref{a15b}). It corresponds to a dynamically generated tree
level gluon mass parameter. At higher orders, quantum polarization
effects will affect the value of that mass parameter. In the absence
of the ghost condensation, this was discussed in
\cite{Browne:2004mk,Gracey:2004bk}.

\noindent For the ghost propagator corresponding to the Overhauser
vacuum given in eq.(\ref{b2}), we have \cite{Dudal:2003dp}
\begin{eqnarray}
\left\langle \overline{c}^{3}(k)c^{3}(-k)\right\rangle &=&\frac{1}{k^{2}\;}%
\;,  \nonumber \\
\left\langle \overline{c}^{\alpha }(k)c^{\beta }(-k)\right\rangle &=&\frac{%
\delta ^{\alpha \beta }k^{2}-\omega \varepsilon ^{\alpha \beta }}{%
k^{4}+\omega ^{2}}\;,\;\;\;\;\alpha ,\beta =1,2\;,  \label{c2}
\end{eqnarray}
where $\varepsilon ^{\alpha \beta }$=$\varepsilon ^{3\alpha \beta
}$. Here we see that the behaviour of the ghost propagator is
changed by the presence of the nonvanishing ghost condensate: there
is a clear distinction between the diagonal and off-diagonal part of
the propagator.

\noindent At one-loop level, the relevant interaction vertex which
has to be taken into account in order to evaluate $\left( \Pi _{\mu
\nu }^{ab}(p,\omega )\right) _{\mathrm{gh}}$, is the
ghost-antighost-gluon vertex, $\varepsilon ^{abc}\partial _{\mu
}\overline{c}^{a}A_{\mu }^{b}c^{c}$.

\subsection{Evaluation of the off-diagonal component $\left( \Pi _{\mu
\nu }^{\alpha \beta }(p,\omega )\right) _{\mathrm{gh}}$.} Let us
consider the off-diagonal component of the ghost contribution to the
vacuum polarization, given by
\begin{equation}
\left( \Pi _{\mu \nu }^{\alpha \beta }(p,\omega )\right)
_{\mathrm{gh}}=g^{2}\varepsilon ^{\alpha mn}\varepsilon ^{\beta pq}\int \frac{d^{d}k}{%
\left( 2\pi \right) ^{d}}\left( p-k\right) _{\mu }k_{\nu
}\left\langle
\overline{c}^{m}c^{q}\right\rangle _{p-k}\left\langle \overline{c}%
^{p}c^{n}\right\rangle _{k}\;.  \label{c10}
\end{equation}
A little algebra results in
\begin{eqnarray}
\varepsilon ^{\alpha mn}\varepsilon ^{\beta pq}\left\langle \overline{c}%
^{m}c^{q}\right\rangle _{p-k}\left\langle
\overline{c}^{p}c^{n}\right\rangle _{k}  &=&-\frac{1}{\left(
p-k\right) ^{2}}\frac{\left( \delta ^{\alpha \beta
}k^{2}+\omega \varepsilon ^{\alpha \beta }\right) }{k^{4}+\omega ^{2}}-\frac{%
1}{k^{2}}\frac{\left( \delta ^{\alpha \beta }\left( p-k\right)
^{2}-\omega \varepsilon ^{\alpha \beta }\right) }{\left( p-k\right)
^{4}+\omega ^{2}}\;, \nonumber\\\label{c11}
\end{eqnarray}
hence
\begin{eqnarray}
\left( \Pi _{\mu \nu }^{\alpha \beta }(p,\omega )\right)
_{\mathrm{gh}} &=&-g^{2}\int
\frac{d^{d}k}{\left( 2\pi \right) ^{d}}\left( p-k\right) _{\mu }k_{\nu }%
\frac{1}{\left( p-k\right) ^{2}}\frac{\left( \delta ^{\alpha \beta
}k^{2}+\omega \varepsilon ^{\alpha \beta }\right) }{k^{4}+\omega
^{2}}
\nonumber \\
&&-g^{2}\int \frac{d^{d}k}{\left( 2\pi \right) ^{d}}\left(
p-k\right) _{\mu }k_{\nu }\frac{1}{k^{2}}\frac{\left( \delta
^{\alpha \beta }\left( p-k\right) ^{2}-\omega \varepsilon ^{\alpha
\beta }\right) }{\left( p-k\right) ^{4}+\omega ^{2}}\;,
\label{c12}
\end{eqnarray}
Use has been made of the property
\begin{equation}
\varepsilon ^{\alpha \delta }\varepsilon ^{\delta \beta }=-\delta
^{\alpha \beta }\;.  \label{c4}
\end{equation}
In the second integral in eq.({\ref{c12}}), we make the change of
variables
\begin{equation}
k_{\mu }\rightarrow p_{\mu }-k_{\mu }\;,  \label{c13}
\end{equation}
to find that
\begin{eqnarray}
\left( \Pi _{\mu \nu }^{\alpha \beta }(p,\omega )\right)
_{\mathrm{gh}} &=&-g^{2}\delta ^{\alpha \beta }\int \frac{d^{d}k}{\left( 2\pi \right) ^{d}}%
\frac{k^{2}\left( p_{\mu }k_{\nu }+p_{\nu }k_{\mu }-2k_{\mu }k_{\nu
}\;\right) }{\left( p-k\right) ^{2}\left( k^{4}+\omega ^{2}\right) }%
\nonumber\\&&-g^{2}\omega \varepsilon ^{\alpha \beta }\int
\frac{d^{d}k}{\left( 2\pi \right) ^{d}}\frac{\left( p_{\mu }k_{\nu
}-k_{\mu }p_{\nu }\right) }{\left( p-k\right) ^{2}\left(
k^{4}+\omega ^{2}\right) }\;.
  \label{c14}
\end{eqnarray}
Since the last term in eq.({\ref{c14}}) vanishes due to its
antisymmetry, we arrive at
\begin{equation}
\left( \Pi _{\mu \nu }^{\alpha \beta }(p,\omega )\right)
_{\mathrm{gh}}=-g^{2}\delta
^{\alpha \beta }\int \frac{d^{d}k}{\left( 2\pi \right) ^{d}}\frac{%
k^{2}\left( p_{\mu }k_{\nu }+p_{\nu }k_{\mu }-2k_{\mu }k_{\nu }\;\right) }{%
\left( p-k\right) ^{2}\left( k^{4}+\omega ^{2}\right) }\;,
\label{c15}
\end{equation}
which coincides in fact with the expression found in
\cite{Sawayanagi:2003dc}.

\noindent Before calculating $\left( \Pi _{\mu \nu }^{\alpha \beta
}(p,\omega )\right) _{\mathrm{gh}}$, we notice, by making use of the
relation
\begin{equation}
\frac{k^{2}}{k^{4}+\omega ^{2}}=\frac{1}{k^{2}}-\frac{\omega
^{2}}{\left( k^{4}+\omega ^{2}\right) k^{2}}\;,  \label{c16}
\end{equation}
that
\begin{equation}
\left( \Pi _{\mu \nu }^{\alpha \beta }(p,\omega )\right)
_{\mathrm{gh}}=\left( \Pi _{\mu \nu }^{\alpha \beta }(p,0)\right)
_{\mathrm{gh}}+\omega ^{2}J_{\mu \nu }^{\alpha \beta }(p,\omega )\;,
\label{c17}
\end{equation}
where
\begin{equation}
\left( \Pi _{\mu \nu }^{\alpha \beta }(p,0)\right)
_{\mathrm{gh}}=-g^{2}\delta ^{\alpha \beta }\int
\frac{d^{d}k}{\left( 2\pi \right) ^{d}}\frac{\left( p_{\mu }k_{\nu
}+p_{\nu }k_{\mu }-2k_{\mu }k_{\nu }\;\right) }{\left( p-k\right)
^{2}k^{2}}\;, \label{c18}
\end{equation}
and
\begin{equation}
J_{\mu \nu }^{\alpha \beta }(p,\omega )=g^{2}\delta ^{\alpha \beta
}\int \frac{d^{d}k}{\left( 2\pi \right) ^{d}}\frac{\left( p_{\mu
}k_{\nu }+p_{\nu }k_{\mu }-2k_{\mu }k_{\nu }\;\right) }{\left(
p-k\right) ^{2}\left( k^{4}+\omega ^{2}\right) k^{2}}\;.
\label{c19}
\end{equation}
One observes that only the term $\left( \Pi _{\mu \nu }^{\alpha
\beta }(p,0)\right) _{\mathrm{gh}}$ is ultraviolet divergent, while
the $\omega $-dependent part $J_{\mu \nu }^{\alpha \beta }(p,\omega
)$ is convergent by power counting. The divergent part of $\left(
\Pi _{\mu \nu }^{\alpha \beta }(p,\omega )\right) _{\mathrm{gh}}$ is
thus the same as if it were computed in the absence of the ghost
condensation.

\noindent To calculate explicitly $\left( \Pi _{\mu \nu }^{\alpha
\beta }(p,\omega )\right) _{\mathrm{gh}}$, we shall concentrate on
\begin{equation}\label{calcul2}
    \pi_{\mu\nu}=\int \frac{d^{d}k}{\left( 2\pi \right) ^{d}}\frac{%
k^{2}\left( p_{\mu }k_{\nu }+p_{\nu }k_{\mu }-2k_{\mu }k_{\nu }\;\right) }{%
\left( p-k\right) ^{2}\left( k^{4}+\omega ^{2}\right) }\;.
\end{equation}
Using
\begin{equation}\label{calcul3}
    \frac{k^2}{k^4+\omega^2}=\frac{1}{2}\left(\frac{1}{k^2+i\omega}+\frac{1}{k^2-i\omega}\right)\;,
\end{equation}
it follows that
\begin{equation}\label{calcul4}
    \pi_{\mu\nu}=\frac{1}{2}\int \frac{d^{d}k}{\left( 2\pi \right) ^{d}}\left(\frac{1}{k^2+i\omega}\right)
    \frac{p_{\mu }k_{\nu }-k_{\mu }k_{\nu } }{\left( p-k\right)
    ^{2}}+\frac{1}{2}\int \frac{d^{d}k}{\left( 2\pi \right) ^{d}}\left(\frac{1}{k^2+i\omega}\right)
    \frac{p_{\nu }k_{\mu }-k_{\mu }k_{\nu } }{\left( p-k\right)
    ^{2}}+(\omega\rightarrow-\omega)\;.
\end{equation}
Setting
\begin{equation}\label{calcul5bis}
\tilde{\pi}_{\mu\nu}=\int \frac{d^{d}k}{\left( 2\pi \right)
^{d}}\left(\frac{1}{k^2+i\omega}\right)
    \frac{p_{\mu }k_{\nu }-k_{\mu }k_{\nu } }{\left( p-k\right)
    ^{2}}\;,
\end{equation}
then clearly
\begin{equation}\label{calcul4bis}
    \pi_{\mu\nu}=\frac{1}{2}\tilde{\pi}_{\mu\nu}+\frac{1}{2}\tilde{\pi}_{\nu\mu}+(\omega\rightarrow-\omega)\;.
\end{equation}
Introducing a Feynman parameter and employing dimensional
regularization, one shall find after some calculation that
\footnote{The property that $\frac{1}{2
i}\ln\frac{1+ix}{1-ix}=\arctan x$ was employed.} for
eq.(\ref{calcul4bis})
\begin{eqnarray}
  \pi_{\mu\nu}&=& \frac{1}{16\pi^2}\left[\frac{\delta_{\mu\nu}}{d}
  \left(\frac{2}{3}\frac{\omega^3}{p^4}\arctan\frac{p^2}{\omega}+\frac{1}{9}\left(\frac{12p^2}{\varepsilon}+13p^2-\frac{6\omega^2}{p^2}\right)\right.\right.
  \nonumber\\ &-&\left.\left.\frac{p^2}{3}\ln\frac{p^4+\omega^2}{\omu^4}+2\omega\arctan\frac{\omega}{p^2}
  -\frac{\omega^2}{p^2}\ln\frac{\omega^2}{p^4+\omega^2}\right)\right.\nonumber\\
  &+&\left.p_\mu p_\nu\left(-\frac{2}{3}\frac{\omega^3}{p^6}\arctan\frac{p^2}{\omega^2}
  +\frac{1}{18}\left(\frac{12}{\varepsilon}+10+12\frac{\omega^2}{p^4}\right)\right.\right.\nonumber\\
  &-&\left.\left.\frac{1}{6}\ln\frac{p^4+\omega^2}{\omu^4}+\frac{1}{2}\frac{\omega^2}{p^4}\ln\frac{\omega^2}{p^4+\omega^2}\right)\right]\;.
\end{eqnarray}
As such, we obtain at last
\begin{eqnarray}
  \left( \Pi _{\mu
\nu }^{\alpha \beta }(p,\omega )\right) _{\mathrm{gh}}&=&
\frac{-g^2\delta^{\alpha\beta}}{16\pi^2}\left[\frac{\delta_{\mu\nu}}{4}
  \left(\frac{2}{3}\frac{\omega^3}{p^4}\arctan\frac{p^2}{\omega}+\frac{1}{9}\left(\frac{12p^2}{\varepsilon}+13p^2-\frac{6\omega^2}{p^2}+3p^2\right)\right.\right.
  \nonumber\\ &-&\left.\left.\frac{p^2}{3}\ln\frac{p^4+\omega^2}{\omu^4}+2\omega\arctan\frac{\omega}{p^2}
  -\frac{\omega^2}{p^2}\ln\frac{\omega^2}{p^4+\omega^2}\right)\right.\nonumber\\
  &+&\left.p_\mu p_\nu\left(-\frac{2}{3}\frac{\omega^3}{p^6}\arctan\frac{p^2}{\omega^2}
  +\frac{1}{18}\left(\frac{12}{\varepsilon}+10+12\frac{\omega^2}{p^4}\right)\right.\right.\nonumber\\
  &-&\left.\left.\frac{1}{6}\ln\frac{p^4+\omega^2}{\omu^4}+\frac{1}{2}\frac{\omega^2}{p^4}\ln\frac{\omega^2}{p^4+\omega^2}\right)\right]\;.\label{calculend}
\end{eqnarray}
According to the analysis of the Slavnov-Taylor identities, the
contribution to the vacuum polarization, that is induced by the
ghost condensation, displays a violation of the transversality, as
it is apparent from the above expression (\ref{calculend}).

\noindent The result (\ref{calculend}) can be compared with that of
\cite{Sawayanagi:2003dc}. In order to do so, we observe that the
$\MSbar$ renormalization was not performed in
\cite{Sawayanagi:2003dc}. However, a careful examination reveals
that our results are in perfect agreement with those of
\cite{Sawayanagi:2003dc}, keeping in mind that
\begin{eqnarray}\label{rep2}
-\gamma+\ln4\pi-\ln
    p^2&\stackrel{\MSbar}{\longrightarrow}&-\ln\frac{p^2}{\omu^2}\equiv-\frac{1}{2}\ln\frac{p^4}{\omu^4}\;,\nonumber\\
    \frac{\pi}{2}-\arctan\frac{1}{x}&=&\arctan x\;,\;\;\;\forall
    x>0\;.
\end{eqnarray}

\subsection{Evaluation of the diagonal component $\left( \Pi _{\mu \nu
}^{33}(p,\omega )\right) _{\mathrm{gh}}$.} Let us take a look at the
diagonal component $\left( \Pi _{\mu \nu }^{33}(p,\omega )\right)
_{\mathrm{gh}}$. From expressions (\ref{c2}) , one has
\begin{eqnarray}
\left( \Pi _{\mu \nu }^{33}(p,\omega )\right) _{\mathrm{gh}}
&=&g^{2}\varepsilon ^{\alpha \beta }\varepsilon ^{\delta \gamma
}\int \frac{d^{d}k}{\left( 2\pi \right) ^{d}}\frac{\left( p-k\right)
_{\mu }k_{\nu }}{\left( \left( p-k\right) ^{4}+\omega ^{2}\right)
\left( k^{4}+\omega ^{2}\right)
}\times\nonumber\\\;\;\;\;\;\;\;\;\;\;\;\;&&(\delta ^{\alpha \gamma
}\left( p-k\right) ^{2}-\omega \varepsilon ^{\alpha \gamma })(\delta
^{\delta \beta }k^{2}-\omega \varepsilon ^{\delta \beta })
\nonumber \\
&=&2g^{2}\int \frac{d^{d}k}{\left( 2\pi \right) ^{d}}\frac{\left( p-k\right)
_{\mu }k_{\nu }}{\left( \left( p-k\right) ^{4}+\omega ^{2}\right) \left(
k^{4}+\omega ^{2}\right) }\left( -\left( p-k\right) ^{2}k^{2}+\omega
^{2}\right)  \label{c3}
\end{eqnarray}
One can check that
\begin{eqnarray}&& \frac{\left( p-k\right)
^{2}k^{2}}{\left( \left( p-k\right) ^{4}+\omega ^{2}\right) \left(
k^{4}+\omega ^{2}\right) }
=\nonumber\\&&\frac{1}{\left( p-k\right) ^{2}k^{2}}-\omega ^{2}\left( \frac{k^{2}}{%
\left( \left( p-k\right) ^{4}+\omega ^{2}\right) \left( p-k\right)
^{2}\left( k^{4}+\omega ^{2}\right) }+\frac{1}{\left( p-k\right)
^{2}\left( k^{4}+\omega ^{2}\right) k^{2}}\right),
 \label{c5}
\end{eqnarray}
leading to
\begin{equation}
\left( \Pi _{\mu \nu }^{33}(p,\omega )\right) _{\mathrm{gh}}=\left(
\Pi _{\mu \nu }^{33}(p,0)\right) _{\mathrm{gh}}+\omega ^{2}J_{\mu
\nu }^{33}(p,\omega )\;, \label{c6}
\end{equation}
where
\begin{equation}
\left( \Pi _{\mu \nu }^{33}(p,0)\right) _{\mathrm{gh}}=-2g^{2}\int \frac{d^{d}k}{%
\left( 2\pi \right) ^{d}}\frac{\left( p-k\right) _{\mu }k_{\nu }}{\left(
p-k\right) ^{2}k^{2}}\;,  \label{c7}
\end{equation}
and
\begin{eqnarray}
J_{\mu \nu }^{33}(p,\omega ) &=&2g^{2}\int \frac{d^{d}k}{\left( 2\pi
\right) ^{d}}\frac{\left( p-k\right) _{\mu }k_{\nu }}{\left(
k^{4}+\omega ^{2}\right) }\times\nonumber\\&&\left( \frac{1}{\left(
\left( p-k\right) ^{4}+\omega ^{2}\right) }+\frac{k^{2}}{\left(
\left( p-k\right) ^{4}+\omega ^{2}\right) \left( p-k\right)
^{2}}+\frac{1}{\left( p-k\right) ^{2}k^{2}}\right) \;. \nonumber \\
\label{c9}
\end{eqnarray}
Again, we observe that only the term $\left( \Pi _{\mu \nu
}^{33}(p,0)\right) _{\mathrm{gh}}$ is ultraviolet divergent, while
the $\omega $-dependent part $J_{\mu \nu }^{33}(p,\omega )$ is
convergent by power counting. The divergent part of $\left( \Pi
_{\mu \nu }^{33}(p,\omega )\right) _{\mathrm{gh}}$ is thus the same
as if it were computed in the absence of the ghost condensation.

\noindent Unlike in the off-diagonal case, we shall not determine
the diagonal part of the polarization tensor,
$\left(\Pi_{\mu\nu}^{33}(p,\omega)\right)_{\mathrm{gh}}$, for
general incoming momentum $p$. To obtain the full expression for
$\left(\Pi_{\mu\nu}^{33}(p,\omega)\right)_{\mathrm{gh}}$, very
tedious calculations would be required. To illustrate how
complicated things can become, we have collected some more details
in the Appendix A.

\sect{Mass splitting.} We now come to another consequence due to the
presence of the ghost and gluon condensation. We recall once more
that the condensate $\left\langle A_\mu^2\right\rangle$ gives rise
to an effective dynamical mass in the gauge fixed action
\cite{Dudal:2003by,Verschelde:2001ia}, as it is apparent from
eqns.(\ref{a1}) and from the propagator in eq.(\ref{c1}). In the
absence of the ghost condensation, this mass is the same for all
colors. In the presence of the ghost condensate
$\left\langle\varepsilon^{3bc}\oc^b c^c\right\rangle$, the
interesting phenomenon of the splitting of the diagonal and
off-diagonal gluon masses takes place, due to quantum corrections
induced by the vacuum polarization.

\subsection{Identification of the mass term.}
As we have seen in the previous section, the $\omega$-dependent part $%
J_{\mu \nu }^{ab}(p,\omega )$ of the ghost contribution to the
vacuum polarization is free from ultraviolet divergences.
Furthermore, according to eqns.(\ref{c19}) and (\ref{c9}), $J_{\mu
\nu }^{ab}(p,\omega )$ attains a finite value at $p=0$. This allows
us to interpret $J_{\mu \nu }^{ab}(0,\omega ) $ as a contribution to
the gluon mass in the effective action. Consider in fact the
one-loop two-point function part of the $1PI$ effective action,
namely
\begin{equation}
\frac{1}{2}\int \frac{d^{4}p}{\left( 2\pi \right) ^{4}}A_{\mu
}^{a}(p)\Pi _{\mu \nu }^{ab}A_{\nu }^{b}(-p)\;,  \label{d1}
\end{equation}
where $\Pi _{\mu \nu }^{ab}$ stands for the complete one-loop vacuum
polarization. Let us consider, in particular, the $\omega$-dependent
part of the ghost contribution
\begin{equation}
\frac{1}{2}\int \frac{d^{4}p}{\left( 2\pi \right) ^{4}}A_{\mu
}^{a}(p)J_{\mu \nu }^{ab}(p,\omega )A_{\nu }^{b}(-p)\;,  \label{d2}
\end{equation}
which can be rewritten as
\begin{equation}
\frac{1}{2}\int \frac{d^{4}p}{\left( 2\pi \right) ^{4}}A_{\mu
}^{a}(p)\left( J_{\mu \nu }^{ab}(p,\omega )-J_{\mu \nu
}^{ab}(0,\omega )\right) A_{\nu }^{b}(-p)+\frac{1}{2}\int
\frac{d^{4}p}{\left( 2\pi \right) ^{4}}A_{\mu }^{a}(p)J_{\mu \nu
}^{ab}(0,\omega )A_{\nu }^{b}(-p)\;.  \label{d3}
\end{equation}
The second term in expression $\left( {\ref{d3}}\right) $ is
interpreted as an induced mass. In the next section, we shall see
that this term will be responsible for the splitting of the masses
of the diagonal and off-diagonal components of the gluon field.

\noindent To avoid any confusion, we mean with mass thus the $1PI$
effective mass obtained from the vacuum polarization at zero
momentum.

\noindent One could also study the pole mass. In the absence of the
ghost condensation, the pole of the (Euclidean) gluon propagator was
calculated in the LCO formalism in
\cite{Browne:2004mk,Gracey:2004bk}. In principle, a study along the
lines of \cite{Browne:2004mk,Gracey:2004bk} might be done also in
the present case, but we would need knowledge of the polarization
tensor at nonvanishing momentum. We mention again that, in case of
$\left( \Pi _{\mu \nu }^{33}(p,\omega )\right) _{\mathrm{gh}}$, this
is a highly complicated task (see Appendix A). A determination of
the pole mass might be useful in order to facilitate a numerical
comparison with other values in the unsplitted case, but this is
beyond the aim of the current paper.

\subsubsection{Evaluation of $\left( \Pi _{\mu \nu }^{ab}(0)\right)
_{\mathrm{gh}}$.} For the one-loop ghost contribution to the vacuum
polarization at zero momentum we get
\begin{equation}
\left( \Pi _{\mu \nu }^{ab}(0)\right)
_{\mathrm{gh}}=\frac{1}{d}\delta _{\mu \nu }\left( \Pi _{\rho \rho
}^{ab}(0)\right) _{\mathrm{gh}}\;,  \label{pv}
\end{equation}
with
\begin{equation}
\left( \Pi _{\rho \rho }^{ab}(0)\right)
_{\mathrm{gh}}=-g^{2}\varepsilon
^{amn}\varepsilon ^{bpq}\int \frac{d^{d}k}{\left( 2\pi \right) ^{d}}%
k^{2}\left\langle \overline{c}^{m}c^{q}\right\rangle
_{k}\left\langle \overline{c}^{p}c^{n}\right\rangle _{k}\;.
\label{v1}
\end{equation}

\subsubsection{Evaluation of the off-diagonal components $\left( \Pi
_{\mu \nu }^{\alpha \beta }(0)\right) _{\mathrm{gh}}$.}
Let us begin by evaluating the off-diagonal components\footnote{%
The mixed component $\left( \Pi _{\rho \rho }^{3 \beta }(0)\right)_{\mathrm{gh}}$, $%
\; $ $\beta =1,2$, is easily seen to vanish.} $\left( \Pi _{\rho
\rho }^{\alpha \beta }(0)\right) _{\mathrm{gh}}$, $\alpha $, $\beta
=1,2$, namely
\begin{equation}
\left( \Pi _{\rho \rho }^{\alpha \beta }(0)\right)
_{\mathrm{gh}}=-g^{2}\varepsilon ^{\alpha mn}\varepsilon ^{\beta
pq}\int \frac{d^{d}k}{\left( 2\pi \right) ^{d}}k^{2}\left\langle
\overline{c}^{m}c^{q}\right\rangle _{k}\left\langle
\overline{c}^{p}c^{n}\right\rangle _{k}\;.  \label{v7}
\end{equation}
From eq.(\ref{c15}) with $p_\mu\equiv0$, one derives
\begin{eqnarray}
\left( \Pi _{\rho \rho }^{\alpha \beta }(0)\right)
_{\mathrm{gh}}&=&2g^{2}\delta
^{\alpha \beta }\int \frac{d^{d}k}{\left( 2\pi \right) ^{d}}\frac{k^{2}}{%
\left( k^{4}+\omega ^{2}\right) \;}=2g^{2}\delta ^{\alpha \beta }\int \frac{%
d^{d}k}{\left( 2\pi \right) ^{d}}\frac{k^{4}}{k^{2}\left(
k^{4}+\omega ^{2}\right)}\;, \label{v8}
\end{eqnarray}
hence
\begin{eqnarray}
\left( \Pi _{\rho \rho }^{\alpha \beta }(0)\right) _{\mathrm{gh}}
&=&2g^{2}\delta ^{\alpha \beta }\int \frac{d^{d}k}{\left( 2\pi \right) ^{d}}%
\left( \frac{1}{k^{2}\;}-\frac{\omega ^{2}}{k^{2}\left( k^{4}+\omega
^{2}\right) \;}\right) \nonumber\\&=&-2\omega ^{2}g^{2}\delta ^{\alpha \beta }\int \frac{%
d^{d}k}{\left( 2\pi \right) ^{d}}\left( \frac{1}{k^{2}\left(
k^{4}+\omega ^{2}\right) \;}\right)  =-\delta ^{\alpha \beta
}\frac{\omega g^{2}}{16\pi }\;. \label{v9}
\end{eqnarray}

\subsubsection{Evaluation of the diagonal component $\left( \Pi _{\mu
\nu }^{33}(0)\right) _{\mathrm{gh}}$.} It remains to evaluate the
diagonal component $\left( \Pi _{\mu \nu }^{33}(0)\right)
_{\mathrm{gh}}$. Setting $p_\mu\equiv0$ in the expression
(\ref{c3}), one derives
\begin{eqnarray}
\left( \Pi _{\rho \rho }^{33}(0)\right) _{\mathrm{gh}} &=&-g^{2}\int \frac{d^{d}k}{%
\left( 2\pi \right) ^{d}}\frac{k^{2}}{\left( k^{4}+\omega ^{2}\right) ^{2}}%
\left( -2k^{4}+2\omega ^{2}\right) \;.
\end{eqnarray}
One can check that this can be rewritten as
\begin{eqnarray}
\left( \Pi _{\rho \rho }^{33}(0)\right) _{\mathrm{gh}}&=&-2g^{2}v^{2}\int \frac{d^{d}k}{\left( 2\pi \right) ^{d}}\left( \frac{1}{%
k^{2}\left( k^{4}+\omega ^{2}\right) }+\frac{2k^{2}}{\left(
k^{4}+\omega ^{2}\right) ^{2}}\right) \;.  \label{v4}
\end{eqnarray}
Observe that the integrals in the left hand side of eq.$\left( {\ref{v4}}%
\right) $ are ultraviolet finite. Making use of
\begin{eqnarray}
\int \frac{d^{4}k}{\left( 2\pi \right) ^{4}}\frac{1}{k^{2}\left(
k^{4}+\omega ^{2}\right) } &=&\frac{1}{32\pi }\frac{1}{\omega }\;,
\nonumber
\\
\int \frac{d^{4}k}{\left( 2\pi \right) ^{4}}\frac{k^{2}}{\left(
k^{4}+\omega ^{2}\right) ^{2}} &=&\frac{1}{64\pi }\frac{1}{\omega
}\;,  \label{v5}
\end{eqnarray}
we obtain
\begin{equation}
\left( \Pi _{\rho \rho }^{33}(0)\right)
_{\mathrm{gh}}=-\frac{g^{2}\omega }{8\pi }\;. \label{v6}
\end{equation}

\noindent We see thus that $\left( \Pi _{\rho \rho }^{33}(0)\right)
_{\mathrm{gh}}\neq \left( \Pi _{\rho \rho }^{\alpha \beta
}(0)\right) _{\mathrm{gh}}$, implying that the ghost
condensate $\left\langle \varepsilon ^{abc}\overline{c}^{b}c^{c}\right%
\rangle $ removes in fact the degeneracy of the gluon mass. Notice,
in particular, that the diagonal component of the vacuum
polarization at zero momentum is twice the off-diagonal part, a fact
already observed in \cite{Sawayanagi:2003dc} in the case of the
Curci-Ferrari gauge, see eq.(28) of \cite{Sawayanagi:2003dc}.

\subsubsection{Determination of the pure gluon component $\left( \Pi
_{\mu \nu }^{ab}(0,\omega )\right) _{\mathrm{gl}}$.} We shall also
need the contributions to the vacuum polarization which are coming
from one-loop diagrams built up without ghosts. Due to the form of
the gluon propagator given in eq.(\ref{c2}), these diagrams give the
same contribution to both diagonal and off-diagonal masses. It is
important to keep in mind that the polarization is not solely
determined from the usual Yang-Mills interactions, by employing the
massive gluon propagator, eq.(\ref{c1})}. We draw attention to the
fact that one should calculate with the action (\ref{a1}) describing
the condensed vacuum, where, next to (\ref{b2}), one has
\begin{eqnarray}
\sigma (x) &=&\sigma _{*}+\widetilde{\sigma }(x)\;, \nonumber
\\
\left\langle \widetilde{\sigma}(x)\right\rangle &=&0\;.
\label{b2bis}
\end{eqnarray}
For example, there will be novel contributions coming from the extra
four-point interaction, that adds to the snail diagram, which is no
longer vanishing when the massive propagator of eq.(\ref{c1}) is
employed. Furthermore, there is a $1PI$ diagram generated from the
$(\wsigma A A)$-vertex, also contributing to $\left( \Pi _{\mu \nu
}^{ab}(0,\omega )\right) _{\mathrm{gl}}$.

\noindent Here, we shall not enter into the details of the
calculation, as the relevant diagrams have already been evaluated in
\cite{Browne:2004mk,Gracey:2004bk}. We shall only quote the result
at $p^2=0$,
\begin{equation}
\left( \Pi _{\mu \nu }^{ab}(0)\right)
_{\mathrm{gl}}=\frac{g^2N}{16\pi^2}m^2\left(-\frac{7}{96}+\frac{17}{16}\ln\frac{m^2}{\omu^2}\right)\delta^{ab}\delta
_{\mu \nu }\;, \label{pvbis}
\end{equation}
where we have dropped already the divergent part in
$\frac{1}{\varepsilon}$, as we are assured that the theory is
renormalizable.

\subsection{Interpretation of the mass splitting.}
As we have seen, the ghost condensate $\left\langle \varepsilon ^{abc}%
\overline{c}^{b}c^{c}\right\rangle $ induces a splitting in the
gluon masses through quantum effects. Also, as observed in
\cite{Dudal:2002xe,Sawayanagi:2003dc}, the contribution of the ghost
condensate to the effective gluon mass is negative. For the
splitting of the gluon masses at one-loop order, we may write
\begin{eqnarray}
m_{\mathrm{diag}}^{2} &=&m^{2}+\delta m^{2}- \frac{g^{2}\omega
}{32\pi }\;,
\nonumber \\
m_{\mathrm{off-diag}}^{2} &=&m^{2}+\delta m^{2}- \frac{g^{2}\omega
}{64\pi }\;,  \label{f1}
\end{eqnarray}
where $\delta m^{2}$ stands for the contribution to the vacuum
polarization at zero momentum following from one-loop diagrams built
up with gluons only. Explicitly, from eq.(\ref{pvbis}),
\begin{eqnarray}
  \delta m^2=\frac{g^2}{16\pi^2}m^2\left(-\frac{7}{48}+\frac{17}{8}\ln\frac{m^2}{\omu^2}\right)\;,\label{deltam}
\end{eqnarray}
where we have set $N=2$.

\noindent We could also use the RG invariance here to sum the
leading logarithms. Contributions $\propto \omega$ shall only be
multiplied by powers of $g^2\ln\frac{\omega}{\omu^2}$ as the limit
$m^2\rightarrow0$ exists, thus terms like
$g^2m^2\ln\frac{\omega}{\omu^2}$ cannot appear. The same holds true
for contributions $\propto m^2$. Similar to what happened in section
3, running quantities will get replaced by their values at the mass
scale $m^2$ or $\omega$. More precisely,
\begin{eqnarray}
m_{\mathrm{diag}}^{2} &=&\overline{m}^{2}+
\frac{\overline{g}^2}{16\pi^2}\overline{m}^2\left(-\frac{7}{48}\right)
- \frac{\widetilde{g}^{2}\widetilde{\omega} }{32\pi }\;,
\nonumber \\
m_{\mathrm{off-diag}}^{2} &=&\overline{m}^{2}+
\frac{\overline{g}^2}{16\pi^2}\overline{m}^2\left(-\frac{7}{48}\right)-
\frac{\widetilde{g}^{2}\widetilde{\omega} }{64\pi }\;,
\label{resultaten}
\end{eqnarray}

\noindent Let us take a look at the numbers. Substituting the
quantities quoted in eq.(\ref{ll16}), we arrive at
\begin{eqnarray}
m_{\mathrm{diag}}^{2} &\approx& 1.66\lms^2\;,
\nonumber \\
m_{\mathrm{off-diag}}^{2} &\approx& 2.34\lms^2\;,  \label{f1bis}
\end{eqnarray}
We notice that
\begin{equation}
m_{\mathrm{diag}}^{2}<m_{\mathrm{off-diag}}^{2}\;,  \label{fr}
\end{equation}
In the MAG, the gap existing between the diagonal and off-diagonal
gluon masses was interpreted as analytical evidence for the Abelian
dominance
\cite{Amemiya:1998jz,Bornyakov:2003ee,Kondo:2000ey,Dudal:2004rx}.
Analogously, we could interpret the result (\ref{f1bis})-(\ref{fr})
as a certain indication that a kind of Abelian dominance might take
place in the Landau gauge too. Of course, the numbers\footnote{We
are unable to provide an estimate in terms of GeV, as, to our
knowledge, an explicit value of $\lms^{N=2;N_f=0}$ is not available
in the existing literature.} in eq.(\ref{f1bis}) should be
interpreted with care: a difference between both masses shows up,
but this difference is far to small to see it as the ultimate proof
of Abelian dominance in the Landau gauge. Even in the MAG, the
existing difference in diagonal and off-diagonal mass is taken only
as an indication. Moreover, it is interesting to observe that in the
Landau gauge, a distinction can arise between the diagonal and the
off-diagonal sectors of the gauge group, albeit the gauge fixing
itself respects the global $SU(2)$ invariance. The MAG already makes
a distinction between the diagonal and off-diagonal part of the
gauge group from the beginning\footnote{As a matter of fact, only
the off-diagonal gauge freedom is fixed. A supplementary condition
has to be imposed on the diagonal part of the gauge freedom, as for
instance an Abelian Landau gauge, as used in
\cite{Fazio:2001rm,Dudal:2004rx,Amemiya:1998jz,Bornyakov:2003ee}.}.
As we have discussed in \cite{Dudal:2004rx}, in the MAG the
condensate $\left\langle\frac{1}{2} A_\mu^a A_\mu^a+\al \oc^a
c^a\right\rangle$, which is the counterpart of $\left\langle
A_\mu^2\right\rangle$ in the Landau gauge, provides a mass only for
off-diagonal gluons, thereby already giving an indication of Abelian
dominance in the low energy region without the inclusion of
$\left\langle\varepsilon^{3bc}\oc^bc^c\right\rangle$. The combined
effect of the ghost condensation as well as of
$\left\langle\frac{1}{2} A_\mu^a A_\mu^a+\al \oc^a c^a\right\rangle$
are currently under investigation in the MAG.

\noindent In the absence of the gluon condensation $\left\langle
A_\mu^2\right\rangle$, thus $m^2\equiv0$, eq.(\ref{f1}) shows that
the diagonal and off-diagonal gluon fields attain an effective mass
which is \emph{tachyonic}. This fact was first observed in
\cite{Dudal:2002xe} in the case of the MAG and later on confirmed in
the Curci-Ferrari gauge \cite{Sawayanagi:2003dc}.

\sect{Conclusion.} We have studied simultaneously the condensation
of the mass dimension two local composite operators $A_\mu^2$ and
$f^{abc}\oc^a c^b$ in the case of $SU(2)$ Yang-Mills gauge theory in
the Landau gauge. This extended the already existing research on the
gluon condensate $\left\langle A_\mu^2\right\rangle$
\cite{Verschelde:2001ia,Dudal:2002pq,Dudal:2003by} and the ghost
condensate $\left\langle \varepsilon^{3bc}\oc^a c^b\right\rangle$
\cite{Lemes:2002rc,Dudal:2003dp}.

\noindent Employing the LCO formalism to construct the one-loop
effective potential, we have shown that both condensates are
dynamically favoured as they lower the vacuum energy. The
renormalizability of the resulting theory has been proven to all
orders by means of the algebraic renormalization technique
\cite{Piguet:1995er}. We also presented a study of some effects
induced by the ghost condensation. We have shown, by analyzing the
Slavnov-Taylor identities in the ghost condensed vacuum and by
explicit one-loop calculations, that the vacuum polarization is no
longer transverse, whereas the gluon propagator is.

\noindent In the LCO formalism, the nonvanishing condensate
$\left\langle A_\mu^2\right\rangle$ gives rise to an effective tree
level gluon mass, thus the lowest order gluon propagator gets
modified, as it is apparent from eq.(\ref{c1}). Likewise, the ghost
condensate $\left\langle \varepsilon^{3bc}\oc^a c^b\right\rangle$
influences the ghost propagator, given in eq.(\ref{c2}). We
determined the one-loop correction to the effective gluon mass and
found that the ghost condensate induces a splitting between the
diagonal and off-diagonal sector. The effective off-diagonal gluon
mass turns out be larger than the diagonal one. This might be
interpreted as a first analytical indication for a possible kind of
Abelian dominance in the Landau gauge, analogously to what was done
in the case of the maximal Abelian gauge in
\cite{Amemiya:1998jz,Bornyakov:2003ee,Kondo:2000ey,Dudal:2004rx}. It
is worth mentioning that, recently, some evidence for the Abelian
dominance in the Landau gauge by lattice numerical simulations was
announced in the works \cite{Suzuki:2004dw,Suzuki:2004uz}, where the
appearance of an Abelian dual Meissner effect in this gauge has been
pointed out. However, in the more recent work
\cite{Chernodub:2005gz} by the same authors and others, this opinion
was changed and it was reported that no evidence for Abelian
dominance is observed in the Landau gauge.

\noindent Finally, we hope that our results could stimulate further
lattice numerical studies of the ghost propagator. It would be very
interesting if, somehow, one would be able to simulate the
Overhauser vacuum (\ref{b2}). This could allow one to investigate
the diagonal and off-diagonal part of ghost propagator, which turns
out to be affected by the ghost condensation, see eqns.(\ref{c2}).
At the time of finishing this work, we have been informed about the
results appearing in \cite{Cucchierighost}. In this paper
\cite{Cucchierighost}, a numerical study of the ghost condensation
in the Overhauser channel for $SU(2)$ lattice gauge theory in the
Landau gauge was performed. The data was fit to the theoretical
prediction given in eq.(\ref{c2}), and assuming a small value of the
ghost condensate, the fitted power law behaviour tended to be $\sim
p^{-4}$, in accordance with the theoretical prediction following
from eq.(\ref{c2}). This is promising as it is first numerical
indication that a ghost condensation might occur in the Landau
gauge, although to obtain more conclusive results, simulations at
larger physical volumes will be certainly necessary.

\section*{Acknowledgments.}
The Conselho Nacional de Desenvolvimento Cient\'{\i}fico e
Tecnol\'{o}gico (CNPq-Brazil), the Faperj, Funda{\c{c}}{\~{a}}o de
Amparo {\`{a}} Pesquisa do Estado do Rio de Janeiro, the SR2-UERJ
and the Coordena{\c{c}}{\~{a}}o de Aperfei{\c{c}}oamento de Pessoal
de N{\'\i}vel Superior (CAPES) are gratefully acknowledged for
financial support. The authors would like to thank A.~Cucchieri for
keeping them informed about the work \cite{Cucchierighost}.

\appendix
\section{Appendix: $\left( \Pi _{\mu \nu }^{33}(p,\omega )\right) _{\mathrm{gh}}$. }
\noindent In this Appendix, we shall outline some details concerning
the evaluation of $\left( \Pi _{\mu \nu }^{33}(p,\omega )\right)
_{\mathrm{gh}}$, given in eq.(\ref{c3}).

\noindent We shall concentrate on
\begin{eqnarray}
\pi^\prime_{\mu\nu}&=&\int \frac{d^{d}k}{\left( 2\pi \right)
^{d}}\frac{\left( p-k\right) _{\mu }k_{\nu }}{\left( \left(
p-k\right) ^{4}+\omega ^{2}\right) \left( k^{4}+\omega ^{2}\right)
}\left( -\left( p-k\right) ^{2}k^{2}+\omega ^{2}\right)
\;.\label{n1}
\end{eqnarray}
Using the decomposition (\ref{calcul3}) twice, one can write
\begin{eqnarray}
\pi^\prime_{\mu\nu}&=&\frac{1}{4}\int \frac{d^d
k}{(2\pi)^d}\frac{\left( p-k\right) _{\mu }k_{\nu
}}{k^2(p-k)^2}\left(
-\left( p-k\right)^{2}k^{2}+\omega ^{2}\right) \nonumber\\
&&\times\left(\frac{1}{(p-k)^2+i\omega}\frac{1}{k^2+i\omega}+\frac{1}{(p-k)^2+i\omega}\frac{1}{k^2-i\omega}\right)+(\omega\rightarrow-\omega)\;,\label{n2}
\end{eqnarray}
meaning that we must calculate
\begin{eqnarray}
\widetilde{\pi}^\prime_{\mu\nu}&=&\omega^2\int \frac{d^d
k}{(2\pi)^d}\frac{\left( p-k\right) _{\mu }k_{\nu
}}{k^2(p-k)^2}\left(\frac{1}{(p-k)^2+i\omega}\frac{1}{k^2+i\omega}+\frac{1}{(p-k)^2+i\omega}\frac{1}{k^2-i\omega}\right)
\nonumber\\
&-&\int \frac{d^d k}{(2\pi)^d}\left( p-k\right) _{\mu }k_{\nu
}\left(\frac{1}{(p-k)^2+i\omega}\frac{1}{k^2+i\omega}+\frac{1}{(p-k)^2+i\omega}\frac{1}{k^2-i\omega}\right)
 \;.\label{n3}
\end{eqnarray}
Therefore, let us study the basic types of integrals we shall need
to perform the full calculation, being
\begin{eqnarray}
  I_{\mu\nu}^\pm&=&\int \frac{d^d
k}{(2\pi)^d}\frac{\left( p-k\right) _{\mu }k_{\nu
}}{k^2(p-k)^2}\left(\frac{1}{(p-k)^2+i\omega}\frac{1}{k^2\pm
i\omega}\right)
\label{t1}\;,\\
 J_{\mu\nu}^\pm  &=&\int \frac{d^d
k}{(2\pi)^d}\left( p-k\right) _{\mu }k_{\nu }\left(\frac{1}{(p-k)^2+
i\omega}\frac{1}{k^2\pm i\omega}\right)\label{t2}\;.
\end{eqnarray}
Let us begin with $I_{\mu\nu}^\pm$. We shall have to employ the
generalized Feynman trick,
\begin{equation}\label{trick}
    \frac{1}{ABCD}=\int_0^1
    dxdydz\frac{6}{\left[xA+yB+zC+(1-x-y-z)D\right]^4}\;,
\end{equation}
which leads to
\begin{eqnarray}
  I_{\mu\nu}^\pm&=&6\int_0^1dxdydz\int \frac{d^d
k}{(2\pi)^d}\frac{p_\mu k_\nu-k_\mu k_\nu}{\left[yp^2-2yp\cdot
k+zp^2-2zp\cdot
k+zi\omega+k^2\pm(1-x-y-z)i\omega\right]^4}\;.\nonumber\\
\end{eqnarray}
The substitution
\begin{equation}\label{subst2}
    K=k-yp-zp\;,
\end{equation}
allows to conclude that
\begin{eqnarray}\label{n5}
  I_{\mu\nu}^\pm&=&6\int_0^1dxdydz\int \frac{d^d
K}{(2\pi)^d}\frac{-\frac{\delta_{\mu\nu}}{d}K^2+p_\mu
p_\nu(y+z)\left(1-(y+z)\right)}{\left[K^2+\Delta^\pm\right]^4}\;,
\end{eqnarray}
where
\begin{equation}\label{n4}
    \Delta^\pm=p^2(-y^2-z^2-2yz+y+z)+zi\omega\pm(1-x-y-z)i\omega\;.
\end{equation}
Both $K$-integrations showing up are finite, and can be directly
computed without any regularization, leading to
\begin{eqnarray}\label{n6}
  I_{\mu\nu}^\pm&=&\int_0^1dxdydz\left(-\frac{\delta_{\mu\nu}}{32\pi^2}\frac{1}{\Delta^\pm}+\frac{p_\mu
  p_\nu}{16\pi^2}(y+z)\left(1-(y+z)\right)\frac{1}{(\Delta^\pm)^2}\right)\;.
\end{eqnarray}
From the previous expression, it might be clear that the triple
integral in $(x,y,z)$ is far from being trivial and would give rise
to a very complicated final result. We shall not attempt to
calculate it any further.

\noindent As a check of the computation of $\left( \Pi _{\mu \nu
}^{33}(p,\omega )\right) _{\mathrm{gh}}$, we could determine its
pole structure and compare it with the output determined by the
symbolic language {\sc Form} \cite{Vermaseren:2000nd}, in which case
the integrals were calculated by expanding them in the external
momentum $p$.  The $\frac{1}{\varepsilon}$ part is completely
determined by the integrations of the type $J_{\mu\nu}$, as
$I_{\mu\nu}$ is finite. We introduce once more a Feynman parameter
to write
\begin{equation}\label{n7}
    J_{\mu\nu}^\pm=\int_0^1dx\int \frac{d^d
k}{(2\pi)^d}\frac{p_\mu k_\nu-k_\mu k_\nu}{(xp^2-2xp\cdot
k+k^2+xi\omega\pm(1-x)i\omega)^2}\;,
\end{equation}
Substituting $K=k-xp$, one finds
\begin{equation}\label{n8}
    J_{\mu\nu}^\pm=\frac{1}{16\pi^2}\int_0^1dx\left[-\delta^\pm\frac{\delta_{\mu\nu}}{d}\left(-\frac{4}{\varepsilon}+2\ln\frac{\delta^\pm}{\omu^2}-1\right)+x(1-x)p_\mu
    p_\nu\left(\frac{2}{\varepsilon}-\ln\frac{\delta^\pm}{\omu^2}\right)\right]\;,
\end{equation}
after the necessary simplifications, where we defined
\begin{equation}\label{n8}
    \delta^{\pm}=-x^2p^2+xp^2+xi\omega\pm(1-x)i\omega\;.
\end{equation}
We remind that
\begin{equation}\label{n9}
    \left( \Pi _{\mu \nu }^{33}(p,\omega )\right) _{\mathrm{gh}}=\frac{g^2}{2}\left[(I_{\mu\nu}^++I_{\mu\nu}^-+(\omega\rightarrow-\omega))
    +(J_{\mu\nu}^++J_{\mu\nu}^-+(\omega\rightarrow-\omega))\right]\;,
\end{equation}
while simplifying the $J_{\mu\nu}^\pm$-part leads to
\begin{eqnarray}\label{n10}
    \left( \Pi _{\mu \nu }^{33}(p,\omega )\right) _{\mathrm{gh}}&=&\frac{g^2}{2}\left[(I_{\mu\nu}^++I_{\mu\nu}^-+(\omega\rightarrow-\omega))\right]\nonumber\\
    &-&\frac{g^2}{2}\frac{1}{16\pi^2}\left[\frac{\delta_{\mu\nu}}{d}\left(\frac{8p^2}{3\varepsilon}+\textrm{finite}\right)+p_\mu
    p_\nu\left(\frac{4}{3\varepsilon}+\textrm{finite}\right)\right]\;,
\end{eqnarray}
and we conclude that
\begin{eqnarray}
  \left( \Pi _{\mu \nu }^{33}(p,\omega )\right) _{\mathrm{gh}}^\mathrm{div}&=&-\frac{g^2}{16\pi^2}\delta_{\mu\nu}\frac{p^2}{3\varepsilon}-\frac{g^2}{16\pi^2}p_\mu
  p_\nu\frac{2}{3\varepsilon}\;.
  \end{eqnarray}
This result is in accordance with the result obtained by using {\sc
Form}. Let us finally mention that the complete result of $\left(
\Pi _{\mu \nu }^{33}(p,\omega )\right) _{\mathrm{gh}}$, is not very
transparent.

\end{document}